\documentclass[iop]{emulateapj}

\usepackage{natbib}
\usepackage{amsmath}

\newcommand{\msun}{M$_{\sun}$}

\newcommand{\ldl}{$\lambda/{\Delta}{\lambda}$}
\newcommand{\teff}{T$_{eff}$}
\newcommand{\logg}{$\log{g}$}

\newcommand{\vsini}{$v\sin{i}$}
\newcommand{\lbol}{$\log_{10}{L_{bol}/L_{\sun}}$}
\newcommand{\lhalbol}{$\log_{10}{L_{H\alpha}/L_{bol}}$}
\newcommand{\lxlbol}{$\log_{10}{L_{X}/L_{bol}}$}
\newcommand{\lrlbol}{$\log_{10}{{\nu_{pk}}L_{\nu}/L_{bol}}$}
\newcommand{\ujy}{$\mu$Jy}
\newcommand{\kms}{km~s$^{-1}$}

\newcommand{\name}{WISE~J072003.20$-$084651.2}
\newcommand{\namesh}{WISE~J0720$-$0846}

\slugcomment{Submitted to AJ 10 July 2015; Accepted 24 August 2015}

\begin{document}

\title{Radio Emission and Orbital Motion from the Close-Encounter Star-Brown Dwarf Binary WISE~J072003.20$-$084651.2\footnote{Some of the data presented herein were obtained at the W.M. Keck Observatory, which is operated as a scientific partnership among the California Institute of Technology, the University of California and the National Aeronautics and Space Administration. The Observatory was made possible by the generous financial support of the W.M. Keck Foundation.}}

\author{Adam J. Burgasser\altaffilmark{1},
Carl Melis\altaffilmark{1}, 
Jacob Todd\altaffilmark{2}, 
Christopher R.\ Gelino\altaffilmark{3,4},
Gregg Hallinan\altaffilmark{4}
\&
Daniella Bardalez Gagliuffi\altaffilmark{1}
}

\altaffiltext{1}{Center for Astrophysics and Space Science, University of California San Diego, La Jolla, CA, 92093, USA; aburgasser@ucsd.edu}
\altaffiltext{2}{Department of Physics and Astronomy, University of California, Los Angeles, CA 90095-1562, USA}
\altaffiltext{3}{NASA Exoplanet Science Institute, Mail Code 100-22, California Institute of Technology, 770 South Wilson Avenue, Pasadena, CA 91125, USA}
\altaffiltext{4}{Infrared Processing and Analysis Center, MC 100-22, California Institute of Technology, Pasadena, CA 91125, USA}

\begin{abstract}
We report the detection of radio emission and orbital motion from the nearby star-brown dwarf binary WISE~J072003.20$-$084651.2AB. Radio observations across the 4.5--6.5\,GHz band with the Very Large Array identify at the position of the system quiescent emission with a flux density of 15$\pm$3~$\mu$Jy, and a highly-polarized radio source that underwent a 2--3~min burst with peak flux density 300$\pm$90~$\mu$Jy. The latter emission is likely a low-level magnetic flare similar to optical flares previously observed for this source. No outbursts were detected in separate narrow-band H$\alpha$ monitoring observations. We report new high-resolution imaging and spectroscopic observations that confirm the presence of a co-moving T5.5 secondary and provide the first indications of three-dimensional orbital motion. We used these data to revise our estimates for the orbital period ({4.1$^{+2.7}_{-1.3}$~yr}) and tightly constrain the orbital inclination to be nearly edge-on ({93$\fdg$6$^{+1\fdg6}_{-1\fdg4}$}), although robust measures of the component and system masses will require further monitoring. 
The inferred orbital motion does not change the high likelihood that this radio-emitting very low-mass binary made a close pass to the Sun in the past 100~kyr.
\end{abstract}


\section{Introduction}

{\name} (hereafter {\namesh}) is an M9.5 dwarf originally identified by \citet{2014A&A...561A.113S} as a possible new member of the local 8~pc sample. It was previously missed in searches for nearby low-mass dwarfs due to its low Galactic latitude ($b = 2.3\degr$) and small proper motion (121.7$\pm$0.3~mas~yr$^{-1}$). Subsequent follow-up observations by \citet{2014ApJ...783..122K}; \citet[hereafter B15]{2015AJ....149..104B}; and \citet[hereafter I15]{2015A&A...574A..64I} confirmed the late-type nature and proximity of the source (6.0$\pm$1.0~pc), and have identified it as having an unusually high recessional velocity (+83.8~{\kms}), indicative of old disk/thick disk kinematics.  \citet{2015ApJ...800L..17M} have deduced that {\namesh} had one of the closest stellar approaches to the Sun inferred to date, passing within 0.25$^{+0.11}_{-0.07}$~pc over a period of 60,000--85,000 years ago, possibly penetrating the outer Oort Cloud. 

Evidence for a T-type brown dwarf companion to {\namesh} was reported by B15, based {on} both near-infrared spectral analysis and the presence of a candidate resolved source in high-resolution laser guide star adaptive optics (LGSAO) imaging. 
In support of this, I15 reported infrared excess {in} 11 and 22~{\micron} photometry from the Wide-field Infrared Survey Explorer (WISE; \citealt{2010AJ....140.1868W}) that could be attributed to a low-temperature secondary.  
However, the detection of the secondary was marginal due to its close separation (140~mas), large magnitude difference with the primary ($\Delta{H} = 4$), and poor LGSAO correction, and neither B15 nor I15 detected significant radial velocity (RV) variability over overlapping $\sim$3~month timescales. 
Confirming the presence of this putative companion and assessing the degree of its gravitational perturbation on the primary is important for determining an accurate parallax and space motion for {\namesh}. This system is only the second (candidate) binary to be identified among the 14 known late-M dwarfs within 10~pc of the Sun.

Another remarkable trait of {\namesh} reported in B15 and I15 is its weak yet highly variable magnetic emission.  Quiescent H$\alpha$ emission was observed to be at or below typical values for M9--L0 dwarfs, with a relative power of {\lhalbol} $\approx$ $-$5.  
However,  B15 reported the detection of multiple flaring events, both in white-light photometry and H$\alpha$ line emission. These flares were infrequent ($<$1\% effective duty cycle) but nevertheless produced order-of-magnitude variations in nonthermal emission. 
Unfortunately, the X-ray limit of {\lxlbol} $\lesssim$ $-$3.2 reported by I15 does not provide a stringent constraint on the high-energy nonthermal emission from this source.
Overall, {\namesh} appears to be similar in magnetic behavior to the rapidly rotating BRI~0021-0040 \citep{1995AJ....109..762B,1999ApJ...527L.105R}, being weakly active but with occasional strong bursts.   Deducing how these bursts relate to age or rotation, or in the case of {\namesh} interaction with a (putative) close companion, may provide critical clues for understanding the overall decline in optical and X-ray magnetic activity across the M dwarf/L dwarf transition \citep{2000AJ....120.1085G,2000AJ....120..447K,2004AJ....128..426W,2006A&A...448..293S,2014arXiv1410.0014S} and contrary trends in nonthermal radio emission \citep{2006ApJ...648..629B,2012ApJ...746...23M,2013A&A...549A.131A}.

In this article, we report new observations of {\namesh} at optical, near-infrared and radio wavelengths which confirm both the bursting and binary nature of this source. In Section~2 we report the detection of low-level quiescent and bursting radio emission based on data obtained with the Karl G.\ Jansky Very Large Array (hereafter VLA), and limits on H$\alpha$ variability from asynchronous narrow-band photometric monitoring. 
In Section~3 we report new imaging and high resolution spectroscopic observations that confirm the presence of a T5.5 companion, and provide the first indications of orbital motion.  We use these data to make constraints on the orbital configuration of the system in Section~4. Our results are discussed in Section~5.

\section{Magnetic Emission from {\namesh}}

\subsection{Observations}

{\namesh} was observed with the VLA (project code 14B$-$313, PI Burgasser)
in the compact C-configuration (baselines of 0.035$-$3.4\,km) on 2014 November 11 
from UT 08:48:50 to 14:47:48.
The WIDAR correlator was set up for C-band continuum observations
with two basebands, each having eight 128\,MHz sub-bands centered at 5.0\,GHz and 6.0\,GHz, for 
a frequency range of 4.488\,GHz to 6.512\,GHz and a total bandwidth of
$\approx$2\,GHz. Each sub-band had
four polarization products (RR, LL, RL, LR) and sixty-four 2\,MHz channels; 
a 5\,sec dump rate was used.
After observation of our primary
calibration source 3C48 to set the absolute flux scale and measure the
complex bandpass, we conducted a sequence of 4\,min cycles, with 3\,min 
on {\namesh} and 1\,min on the gain calibrator
QSO\,J0730$-$1141. This observational strategy {optimizes} image quality by frequent monitoring (and hence correction) of phase fluctuations.

All data were reduced with the Astronomical Image Processing Software package
(AIPS; \citealt{2003ASSL..285..109G}) following best practices for wide-band radio data reduction.
Radio-frequency interference (RFI) was present throughout the observation, being
especially persistent and strong in the sub-bands covering 5.782\,GHz to 6.512\,GHz, and 
also present at varying levels in several other sub-bands. All data affected by RFI were
removed. In addition, sub-band 0 of each baseband failed to produce robust data, and these measurements were
also removed. After data flagging, our average frequency was
5.27\,GHz and our total bandwidth was reduced to $\approx$1.2\,GHz.

\subsection{Quiescent Emission}

\begin{figure}
\epsscale{1.0}
\plotone{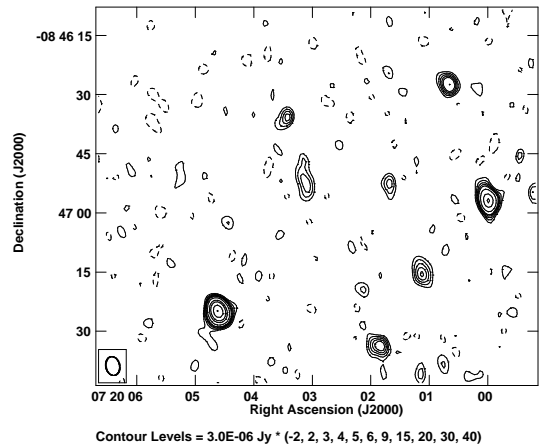}
\caption{VLA image of the {\namesh} field integrated over 4.5--6.5~GHz and over the entire time series on 2014 November 14 (UT).  Contour levels in flux density are labeled at bottom. The beam shape is indicated in the lower left corner and has dimensions of 4$\farcs$8$\times$3$\farcs$5 with position angle of 11$\fdg$5.  Emission from {\namesh} emerges from an extended source spanning $\sim$10$\arcsec$ along a north-south axis, likely arising from combined emission from the target and an unassociated background source.}
\label{fig:vla-image}
\end{figure}

\begin{figure*}
\epsscale{1.1}
\plottwo{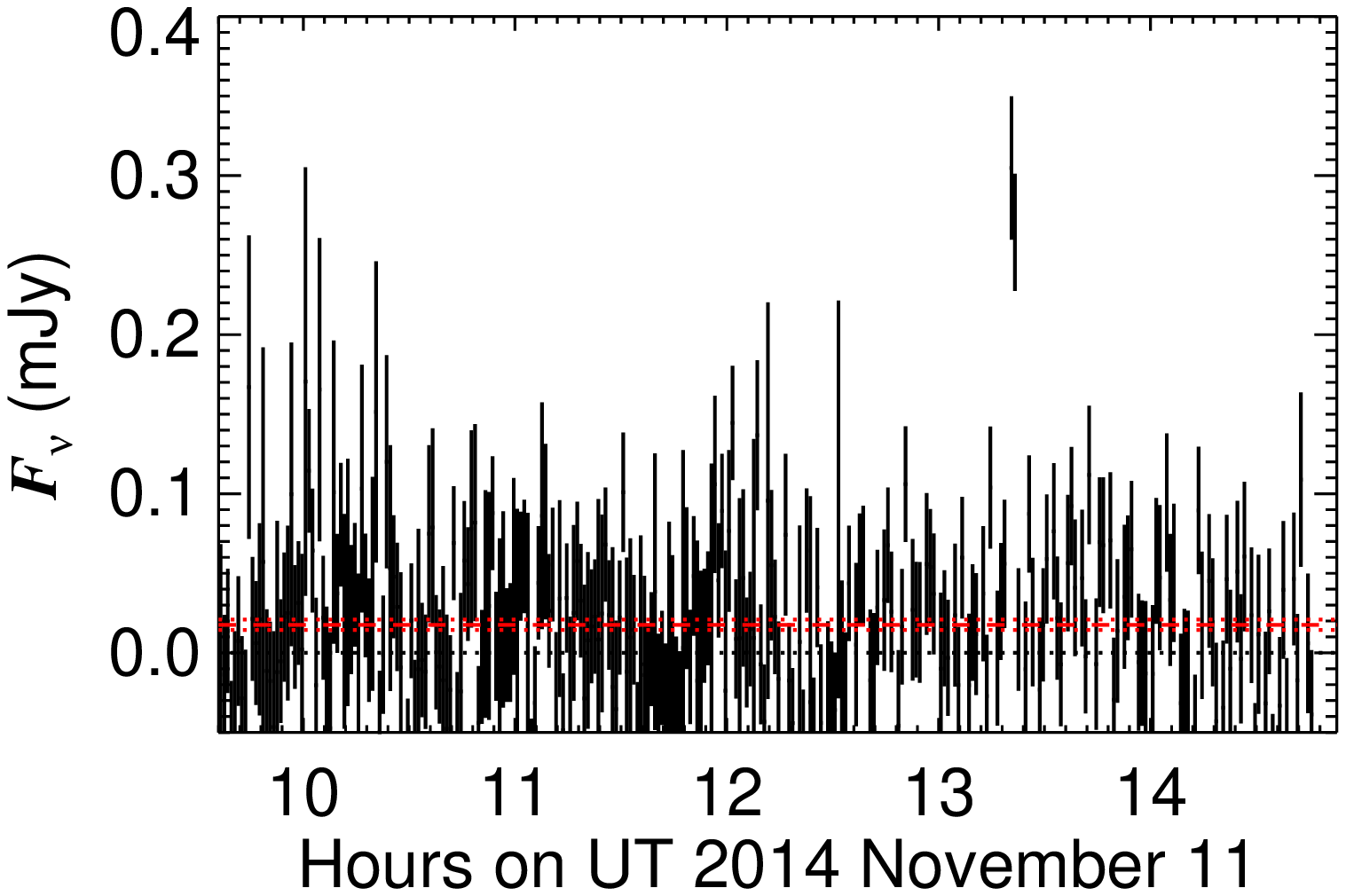}{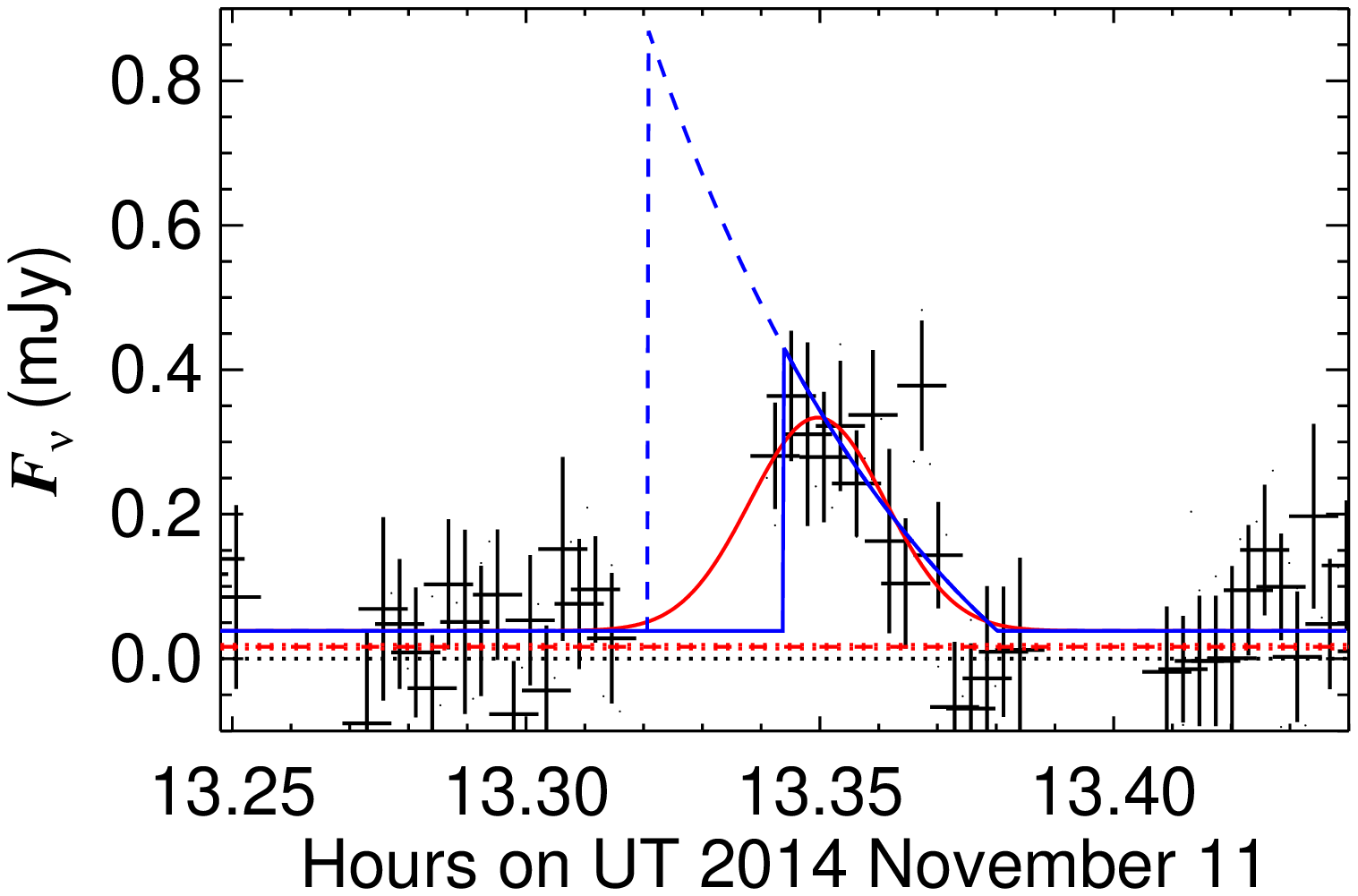}
\caption{(Top) Time series of radio flux from {\namesh} in the 4.5--6.5~GHz band {with 1~min sampling; uncertainties are indicated by error bars.} The red dashed and dotted lines indicate the mean flux level {and 1$\sigma$ uncertainty, after} excluding the bursting emission at UT 13:21.
(Bottom) Close-up view of the burst emission, with data sampled every 10~s. Breaks correspond to observations of the secondary calibrator. The best-fit Gaussian (red solid curve) and exponential decay (blue curves: dashed for peak emission at the start of calibration observation, solid for peak emission at the end) models are overplotted.}
\label{fig:vla-series}
\end{figure*}

The final cleaned and calibrated data were analyzed
by imaging and performing time-series analysis on the $uv$-data. 
Figure \ref{fig:vla-image} shows imaging of the
entire data set (all unflagged times and frequencies)  
at the coordinates of {\namesh}, which reveals a weak, elongated source. 
This is interpretted to be a blend between
weak quiescent emission from {\namesh} and an unrelated faint background
source, each having similar flux densities of $\approx$15\,$\mu$Jy.\footnote{A resolved-object fit to this elongated source yields a total integrated flux density of
28$\pm$8\,$\mu$Jy; the map rms noise level is 2.8\,$\mu$Jy\,bm$^{-1}$} 
Using the AIPS task {\sf DFTPL} and the procedures described in \citet{2009ApJ...691.1128O},
we analyzed the time-series data at the peak flux position, shown in 
Figure~\ref{fig:vla-series}.  This series reveals weak but nonzero emission over the entire course of the observation, with a short burst of emission at UT 13:21:37. Masking out the burst, we measure a mean quiescent flux density of 15$\pm$3~{\ujy} with no evidence of statistically significant variability.\footnote{For 1~minute sampling, we measure $\chi^2$ = 296.0 for 295 degrees of freedom, yielding a $p$-value of 0.46. Note that with the bursting emission included $\chi^2$ = 382.5 and $p$ $<$0.1\%.}
This is the weakest quiescent emission detected for a very low mass dwarf to date, comparable to the quasi-quiescent emission reported for the T6.5 dwarf 2MASS~J10475385+2124234 \citep[16$\pm$5~$\mu$Jy]{2013ApJ...767L..30W}.
The apparent flux density translates into a specific radio luminosity of $\log_{10}{L_{\nu}}$ = (7$\pm$2)$\times$10$^{11}$~erg~s$^{-1}$~Hz$^{-1}$ at the 6.0$\pm$1.0~pc distance of {\namesh}, and {\lrlbol} = $-$8.49$\pm$0.19 assuming {\lbol} = $-$3.60$\pm$0.05 (B15; I15) and $\nu_{pk}$ = 5.28\,GHz.  The specific luminosity is 1--2 orders of magnitude below measurements of previously detected late M and L dwarfs \citep{2013A&A...549A.131A} and the relative luminosity falls below the activity-rotation relation of \citet{2012ApJ...746...23M} given this source's {\vsini} = 8.6$\pm$0.8~{\kms} (Section~3.4).  Indeed, in terms of rotation and radio power, {\namesh} is similar to the M7 {dwarf} VB~8 \citep{1999AJ....118.1369K} and the M8 {dwarf} DENIS~J1048-3956 \citep{2005ApJ...626..486B}, both nearby late M dwarfs with weak but variable magnetic emission and modest rotation rates.  The weakness of the quiescent emission prevents us from determining either its spectral behavior across the 4.4--6.5~GHz band or its polarization.

\subsection{Bursting Emission}

\begin{figure*}
\epsscale{1.0}
\plottwo{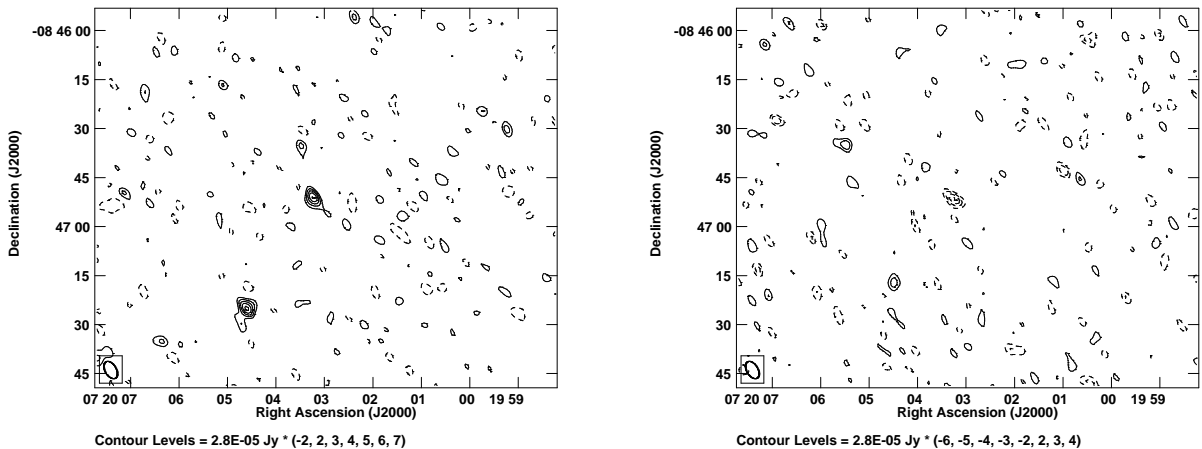}{f3b.eps}
\caption{VLA image of the {\namesh} field in Stokes $I$ (left) and $V$ (right) integrated over 4.5--6.5~GHz and over the period UT 13:19:07 to 13:24:07.  The burst emission seen in the time series data is associated with a point source at the expected position of {\namesh}.  The Stokes $V$ image shows a negative source of comparable brightness indicating nearly 100\% left circular polarization.}
\label{fig:vla-flare-image}
\end{figure*}

A close-up view of the radio burst around UT 13:21 is shown in Figure~\ref{fig:vla-series}. The burst appears to have begun during observations of the secondary calibrator, so our data sample it only at or after its peak emission.  Reimaging of the radio data in the period around this burst shows it to be associated with a bright point source at the coordinates expected for the brown dwarf (Figure~\ref{fig:vla-flare-image}).\footnote{The source identified during the flare
is free of contamination because the instantaneous sensitivity over the small time range imaged
is insufficient to detect the background blend source.}
The burst is highly polarized, with emission in Stokes $I$ of 175$\pm$28~{\ujy} and Stokes $V$ of $-$150$\pm$28~{\ujy}, implying left-handed circular polarization of 84$^{+16}_{-20}$\% (1\,$\sigma$ equivalent uncertainties); i.e., 
consistent with full polarization.
There is no evidence of spectral variation in the burst to within our measurement uncertainties.

We modeled the time series emission as both a rotating spot (``pulse'') feature with a Gaussian profile, and as an exponentially declining flare, $f_{\nu} \propto {\rm e}^{-t/\tau}$ after peak emission. The former yields a peak flux density of 310$\pm$40~$\mu$Jy, $\approx$20 times greater than the quiescent emission, and consistent with a total emitted energy of (1.3$\pm$0.4)$\times$10$^{24}$~erg over the 4.5--6.5~GHz band (assuming a flat spectrum).  The flare model yields a decay time constant of $\tau$ = 3.5$\pm$0.9~min and, depending on whether the burst initiated at the start or end of the calibration period, a total emitted energy of (1--3)$\times$10$^{24}$~erg.  

The burst emission occurs for at most 3~min during our 5~hr of observation of {\namesh}, or 1\% of the total time on-source.
This is consistent with the flaring duty cycle inferred from the aperiodic white light bursts reported in B15.  Thus, we favor an infrequent flare mechanism for this emission as well.
We nevertheless note that the pulse/spot model implies an emission region $\approx$1--2$\degr$ in longitudinal extent (assuming a rotation period of $\approx$14~hr; see Section~4), which is consistent with previous periodic pulse detections \citep{2007ApJ...663L..25H,2015arXiv150206610W}. Longer-term monitoring would be required to distinguish between the flare and pulse hypotheses.

\subsection{H$\alpha$ Monitoring}

\begin{deluxetable*}{lcccccc}
\tablecaption{H$\alpha$ Monitoring Observations\label{tab:halpha}}
\tabletypesize{\small}
\tablewidth{0pt}
\tablehead{
\colhead{UT Date} &
\colhead{MJD} &
\colhead{t$_{int}$} &
\colhead{N$_{obs}$} &
\colhead{UT Start} &
\colhead{Duration} &
\colhead{Mean S/N per}  \\
& &
\colhead{(s)} & & & 
\colhead{(hr)} & \colhead{Measurement} 
}
\startdata
2014 Feb 18 & 56706.26 & 900 & 11 & 4:42:32 & 3.03 & 4 \\
2014 Feb 20 & 56708.24 & 900 & 10 & 4:38:51 & 2.63 & 3 \\
2014 Feb 22 & 56710.24 & 300 & 38 & 3:55:13 & 3.97 & 4 \\
2014 Feb 24 & 56712.28 & 300 & 13 & 4:55:02 & 3.52 & 3 
\enddata
\end{deluxetable*}

{\namesh} was monitored over four nights on 2014 February 18, 20, 22 and 24 (UT) using the facility CCD camera on the 1-meter Nickel telescope at Lick Observatory (Table~\ref{tab:halpha}).  The CCD was configured for 2$\times$2 binning for a pixel scale of 0$\farcs$37~pixel$^{-1}$.  After acquisition and centering at $I$-band, {\namesh} was monitored without dithering through the narrow-band H$\alpha$ filter ($\lambda_c$ = 6557~{\AA}, $\Delta\lambda$ = 15~{\AA}), with integration times of 900~s on February 18 and 20 and 300~s on February 22 and 24.  Total monitoring periods per night spanned 2.63--3.97~hr, but due to overheads and {pauses during occasional clouds} the on-source time {totalled} 7.25~hr over the entire observing run. Bias frames and quartz flat field lamps were also acquired each night for detector calibration.

Data were reduced using standard image reduction techniques. Aperture photometry of {\namesh} and nearby non-saturated stars was measured using a variable aperture scaled to encapsulate 80\% of each source's peak brightness, and an annulus of 63--100~pixels (23$\arcsec$--36$\arcsec$) was used to subtract foreground emission. We did not observe a photometric calibrator during these observations, so no attempt was made to measure absolute H$\alpha$ fluxes. Instead, we used the mean flux of {non-variable} stars in the field of view to compute a reference light curve, and used this to normalize the photometry for {\namesh} over the course of each night.  Uncertainties were dominated by source photometry, of order 20--50\%.

We found no significant flux variations over the course of the four nights of observation.  At the measured noise level, we can rule out bursts  2--3 times above quiescent emission.  B15 reported a single H$\alpha$ flare with a line flux 5 times greater than quiescent emission, and an order of magnitude brighter than the local continuum. Such a flare would have been easily detected in our observations had it occured. Assuming a flare period of $\approx$5~min, we infer a flare indicidence rate of $<$1\%, similar to the white light flare rates reported in B15 and the radio flare reported here. All of these observations reinforce the conclusion that {\namesh} is a weakly active and infrequently bursting source.

\section{Confirming the Multiplicity of {\namesh}}

\subsection{High Resolution Imaging}

\begin{figure*}
\epsscale{0.7}
\plotone{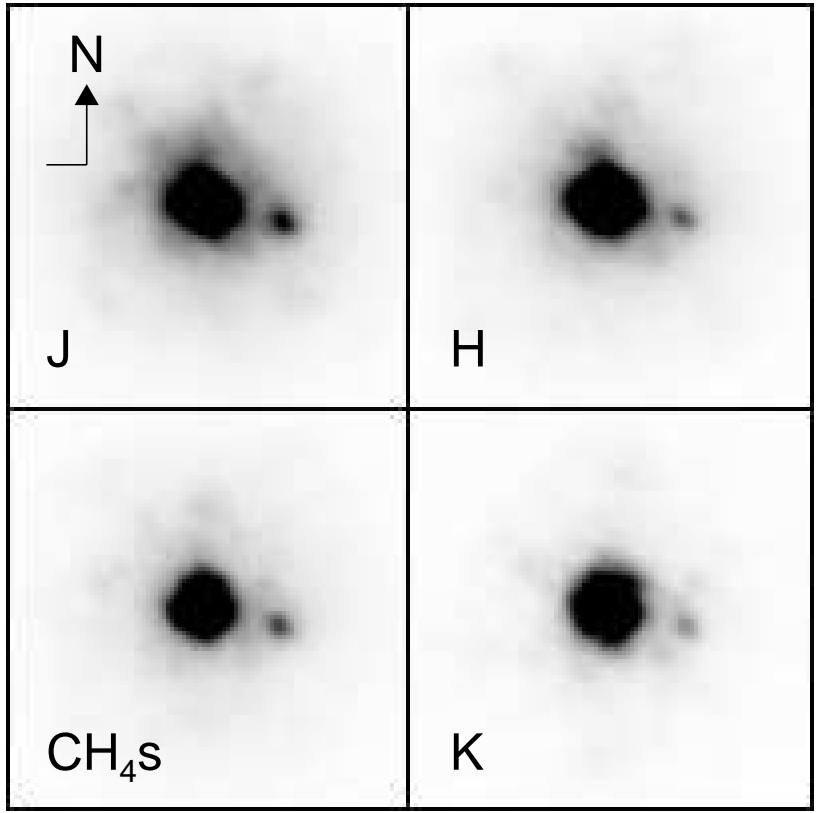}
\caption{Resolved imaging of the {\namesh} system with NIRC2/LGSAO.  Images are aligned with North up and East to the left, and each box displays an angular scale of 1$\arcsec$.  Images are logarithmically flux-scaled to make the faint secondary visible.}
\label{fig:image}
\end{figure*}

{\namesh} was re-observed with the Near-InfraRed Camera 2 (NIRC2) and
LGSAO system \citep{2006PASP..118..310V,2006PASP..118..297W} on the Keck~II 10~m Telescope on 2015 January 11 (UT) in mostly clear and windy conditions with 0$\farcs$8 seeing.  
The narrow field-of-view (FOV) camera was used to obtain dithered observations in 
broad-band MKO\footnote{Mauna Kea Observatories near-infrared filter set \citep{2002PASP..114..169S,2002PASP..114..180T}.} $J$, $H$ and $K_s$ filters, and the medium-band $CH_4s$ filter sampling 1.54--1.65~$\micron$.
The $R=16.8$~mag field star USNO~0812-0137390 was used to correct for tip-tilt aberrations. We achieved better Strehl ratios than observations reported in B15, 10--20\% depending on wavelength. This allowed us to easily resolve the candidate companion reported in that study in all four filters, as shown in Figure~\ref{fig:image}.

\subsection{Common Proper Motion and Orbital Motion}

To extract relative photometry and astrometry, we performed point source function (PSF) fitting of each individual image using a two-dimensional asymmetric Moffat profile optimized to the PSF of the primary component (i.e., with the secondary masked). Measurements are reported in Table~\ref{tab:nirc2}. We find a separation of 197$\pm$3~mas at position angle 256$\fdg$7$\pm$0$\fdg$6, wider than, and at a marginally distinct position angle as, the candidate source previously reported (139$\pm$14~mas at 262$\degr\pm$2$\degr$). As the source detected in these data would have been easily resolved in prior observations, we conclude that statistically significant relative motion has been observed between the two epochs.

\begin{deluxetable}{ll}
\tablecaption{Resolved Photometry and Astrometry of {\namesh} \label{tab:nirc2}}
\tabletypesize{\small}
\tablewidth{0pt}
\tablehead{
\colhead{Parameter} &
\colhead{Value} 
}
\startdata
$\Delta{J}$ & 2.92$\pm$0.07 \\
$\Delta{CH_4s}$ & 3.05$\pm$0.08 \\
$\Delta{H}$ & 3.85$\pm$0.11 \\
$\Delta{K_s}$ & 4.07$\pm$0.20 \\
$\rho$ (mas) & 197$\pm$3 \\
$\rho$ (AU) & 1.19$\pm$0.21 \\
$PA$ ($\degr$) & 256.7$\pm$0.6 \\
Primary SpT\tablenotemark{a} & M9.5$\pm$0.5 \\
Secondary SpT\tablenotemark{a} & T5.5$\pm$0.5 
\enddata
\tablenotetext{a}{Based on spectral template fitting with templates constrained to have the same relative flux scaling as measured in the $JHK_s$ NIRC2 bands.}
\end{deluxetable}

\begin{figure}
\epsscale{1.1}
\plotone{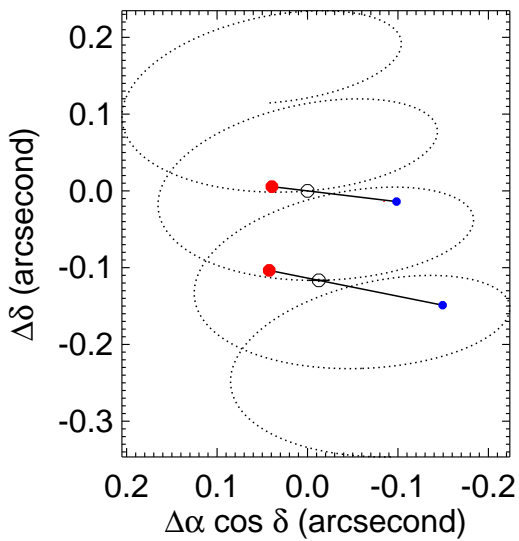}
\caption{Positions of {\namesh}A (red points) and B (blue points) in our 2014 and 2015 LGSAO images relative to center-of-mass astrometric motion (black points and dotted line) based on values from B15 and assuming $q$ = 0.4. The 3$\sigma$ uncertainties on the center-of-mass position at the 2015 epoch {relative to 2014} are indicated: {$\sigma_{\alpha}$ = 4~mas, $\sigma_{\delta}$ = 0.8~mas}. The relative motion of the two sources is consistent with physical association and orbital motion. }
\label{fig:cpm}
\end{figure}

Relative motion can be due to differential motion between two physically unrelated sources or orbital motion in a gravitationally bound binary.  Figure~\ref{fig:cpm} displays the estimated center of mass motion and component positions between our 2014 and 2015 imaging data, assuming systemic astrometry from B15 and $q$ = 0.4 (see below). It is clear that the predominant motion of both sources is co-aligned with the proper motion of the system, particularly in declination; the change in position angle is inconsistent with the secondary being an unmoving background source at the 16$\sigma$ confidence level.  We therefore {determine} that {\namesh} is a co-moving binary system, and identify significant astrometric orbital motion over a one year-period.

\subsection{Improved Determination of the Secondary Classification}

\begin{figure}
\epsscale{1.1}
\plotone{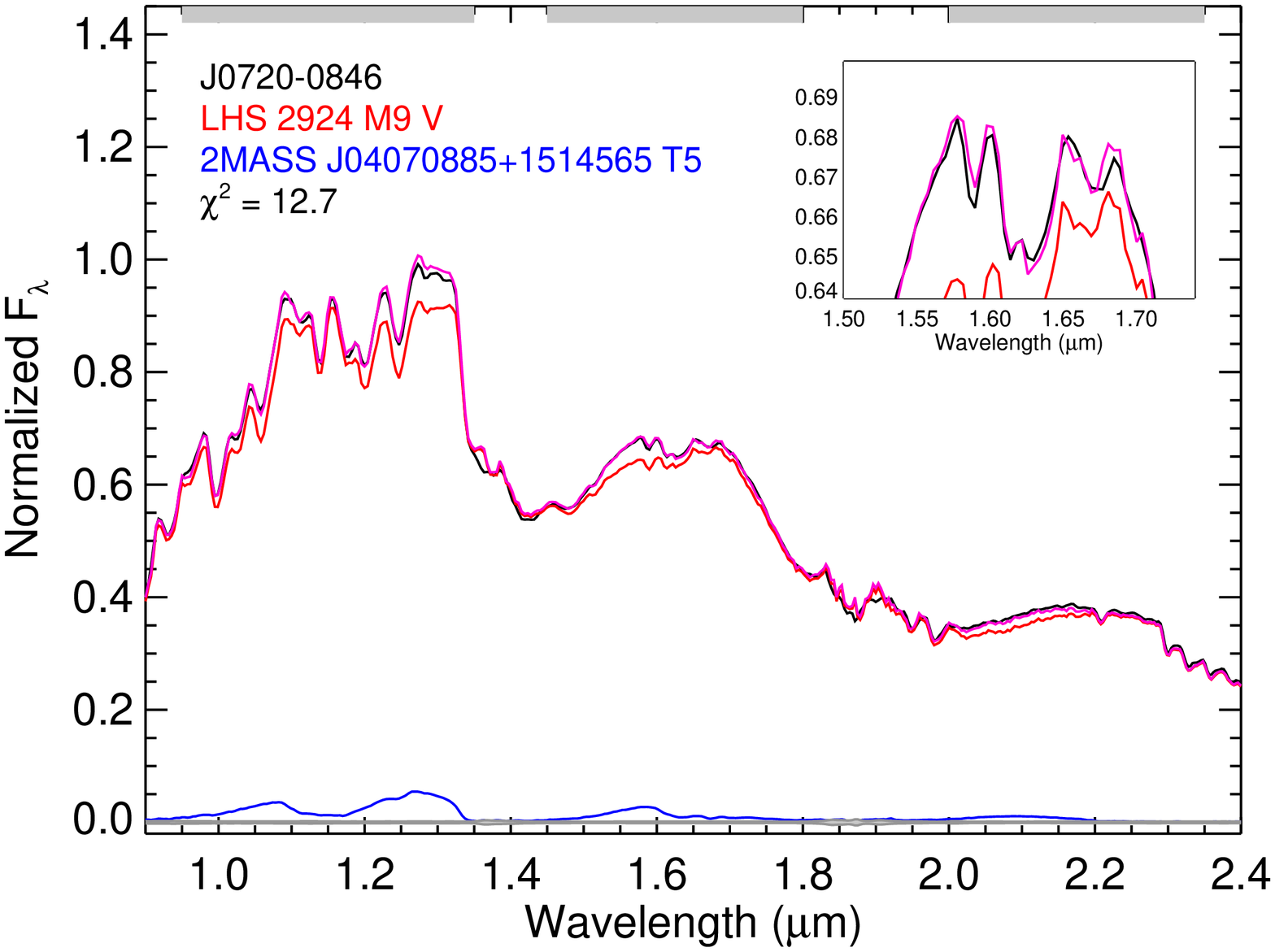}
\caption{Best-fit spectral binary template (purple line) to the combined-light SpeX spectrum of {\namesh} (black line) based on scaling spectral templates to the relative $JHK_s$ photometry measured from NIRC2 observations.  Best-fit primary (red line; {LHS~2924, data from \citealt{2006AJ....131.1007B}}) and secondary templates (blue line, {2MASS~J04070885+1514565, data from \citealt{2004AJ....127.2856B}}) are shown at their relative scaling.  The gray bars at top indicate the regions over which the fitting was done.}
\label{fig:binfit}
\end{figure}

Our new observations provide greatly improved relative photometry of the two components in the four bands measured, and refine the coarse $\Delta{H}$ estimate from B15 (Table~\ref{tab:nirc2}). There is a significant difference in relative brightnesses in the $CH_4s$ and $H$-band filters consistent with strong CH$_4$ absorption in the secondary.  
We used the relative photometry to better constrain the spectral types of the components through spectral template fitting. Following the procedures described in \citet{2011AJ....141...70B}, we combined 512 M7-L1 and 125 T0-T7 spectral templates from the SpeX Prism Library (SPL; \citealt{2014arXiv1406.4887B}), scaled so that that relative spectrophotometry agreed to within 1$\sigma$ of the NIRC2 $JHK_s$ measurements.\footnote{NIRC2 $CH_4s$ photometry was not used for this analysis as the filter profile was unavailable in a digital format (H. Tran, 2015, priv.\ comm.).}  We compared the 16394 binary templates that satisfied these constraints to the combined light SpeX spectra of {\namesh} from both \citet{2014ApJ...783..122K} and B15, following the methods described in the latter paper.  Figure~\ref{fig:binfit} shows the best fitting template to the B15 spectrum. Both analyses yield identical results, with component types M9.5$\pm$0.5 and T5.5$\pm$0.5. The secondary is a half subtype later than, but formally consistent with, the classification reported in B15.

\subsection{High Resolution Spectroscopy}

\begin{figure*}
\epsscale{0.9}
\plotone{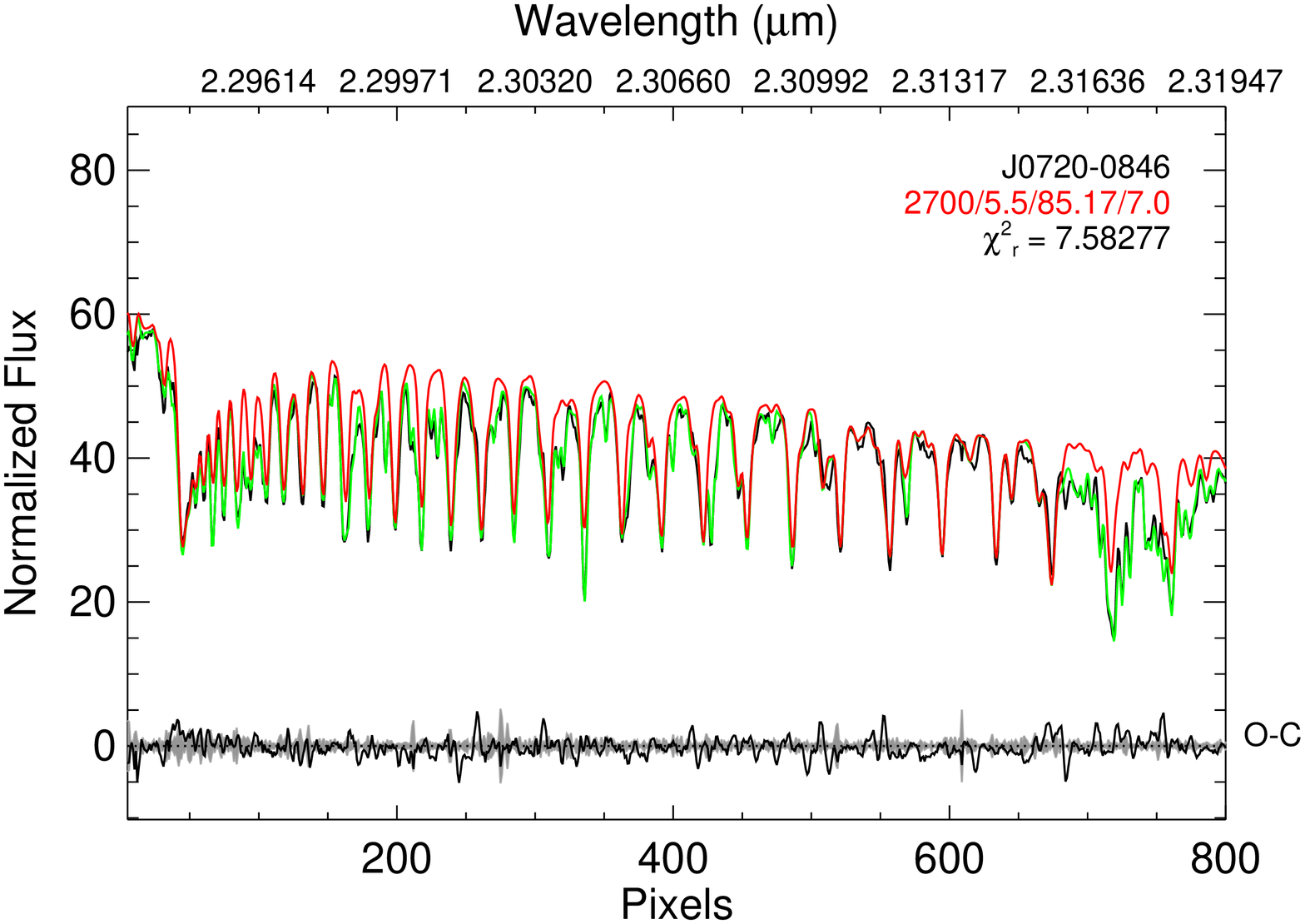} 
\caption{Extracted high-resolution ({\ldl} = 20,000) $K$-band spectrum of {\namesh} obtained with NIRSPEC on UT 2014 December 8 (black line), compared to a best-fit model combining a {{\teff} = 2700~K, {\logg} = 5.5} atmosphere model from \citet[red line]{2012RSPTA.370.2765A} with scaled telluric absorption (green line).  The difference between data and model (O-C)  is shown in black at the bottom of the plot and is dominated by fringing residuals; the {$\pm$1$\sigma$} uncertainty spectrum is indicated in grey.
\label{fig:nirspec}}
\end{figure*}

New high resolution optical and near-infrared spectroscopy of {\namesh} were obtained with 
the Near InfraRed Spectrometer (NIRSPEC; \citealt{2000SPIE.4008.1048M}) on the Keck II telescope on 2014 December 8 (UT), and with
the Hamilton echelle spectrograph \citep{1987PASP...99.1214V} on the Lick Observatory Shane 3~m telescope on 2015 March 10 (UT).
Data were acquired and reduced as described in B15. 
The NIRSPEC spectrum (Figure~\ref{fig:nirspec}) is of high quality (median S/N = 82) and similar to data reported in B15.
The Hamilton data had a somewhat lower signal-to-noise (S/N) than prior observations (S/N $\approx$ 5 at 7500~{\AA}) due to a shorter total integration of 3000~s.

Both spectra were analyzed as described in B15 for radial motion and, for the Hamilton data, H$\alpha$ emission equivalent width (EW).  Despite the low S/N, cross-correlation of the Hamilton spectrum with contemporaneous observations of the RV standard GJ~251 yielded {an} RV = +86.4$\pm$0.5~{\kms}, significantly different than the mean motion reported in B15 (+82.5$\pm$0.4~{\kms}).  Marginal H$\alpha$ emission was observed in these data, with EW = $-$2$\pm$1~{\AA}, consistent with the lowest emission states observed in B15.

For the NIRSPEC data, we re-analyzed the observations reported here and the 2014 observations using an updated Markov Chain Monte Carlo (MCMC) adaptation of the forward-modeling method described in \citet{2010ApJ...723..684B} and B15. We 
used the BT-Settl atmosphere models \citep{2012RSPTA.370.2765A} with updated solar abundance values from \citet{2011SoPh..268..255C} over an effective temperature ({\teff}) range of 1600--2900~K and surface gravity ({\logg}) range of 4.5--5.5 (cgs).  The Solar atlas of \citet{1991aass.book.....L} was used to model telluric absorption features. Table~\ref{tab:nirspec} summarizes the RV and rotational velocities ({\vsini}) inferred from these analyses, while Figure~\ref{fig:nirspec} shows the best-fit model, {with {\teff} = 2700~K and {\logg} = 5.5,} to the most recent NIRSPEC observations.  
The rotational velocity measurements are mutually consistent, with a mean value of {\vsini} = 8.6$\pm$0.4~{\kms}. 
However, like the Hamilton observations, the most recent NIRSPEC RV measurement of +85.2$\pm$0.2~{\kms} is significantly distinct from the mean RV reported in B15 (+83.7$\pm$0.4~{\kms}) and follows the trend of increasing recessional motion.
With these independent measurements, we  conclude that {\namesh}A is an RV variable gravitationally perturbed by its brown dwarf secondary.

\begin{deluxetable}{llcc}
\tablecaption{Radial and Rotational Velocities from NIRSPEC Observations\label{tab:nirspec}}
\tabletypesize{\small}
\tablewidth{0pt}
\tablehead{
\colhead{UT Date} &
\colhead{MJD} &
\colhead{RV} &
\colhead{\vsini} \\
& &
\colhead{({\kms})} &
\colhead{({\kms})} 
}
\startdata
2014 Jan 19 & 56676.00968  & +84.0$\pm$0.3 & 8.8$\pm$0.7 \\
2014 Jan 20 & 56677.00048 &  +83.2$\pm$0.2 & 9.3$\pm$0.8 \\
2014 Mar 10 & 56725.71832 &  +82.9$\pm$0.5 & 6.6$\pm$2.3 \\
2014 Apr 12 & 56758.74056 &  +84.3$\pm$0.4 & 9.9$\pm$1.8 \\
2014 Dec 8 & 56999.00802 &  +85.2$\pm$0.3 & 7.6$\pm$0.8 \\
\hline
Mean & \nodata & +83.8$\pm$0.8\tablenotemark{a} & 8.6$\pm$0.8 
\enddata
\tablenotetext{a}{$\chi^2$ = 35.9 indicates data inconsistent with a constant radial velocity.}
\end{deluxetable}

We note that the best-fitting models for the forward-modeling analyses were consistently in the ranges {\teff} = {2500--2800~K and {\logg} = 5.0--5.5. The temperatures are somewhat higher than those expected for an M9.5 dwarf ({\teff} $\approx$ 2300~K; \citealt{2009ApJ...702..154S}), but the surface gravity is consistent with the lack of Li~I absorption and low-surface gravity spectral features, indicating an age $\gtrsim$1~Gyr.  We verified that lower temperature and lower surface gravity models yielded identical RVs, so this measurement appears to be insensitive to the specific model over the parameter range examined here.  A robust investigation of the atmospheric parameters of this source is deferred to a later study}.

\section{Improved Constraints on the Orbit of {\namesh}AB}

\subsection{Methodology}

With {\namesh}AB verified as a gravitationally bound binary with orbital motion detected in all three spatial dimensions, we can begin to constrain the orbital properties of the system and the physical properties of its components. We adapted the MCMC analysis described in \citet{2012ApJ...757..110B} to include both RV and relative astrometry measurements.  {We employed an orbit} model with nine parameters,
\begin{equation}
\vec{\theta} =  \left(P,a,e,i,\omega,\Omega,M_0,q,V_{COM},d\right)
\end{equation}
where $P$ is the period of the orbit in years, $a$ is the semi-major axis in AU, $e$ is the eccentricity, $i$ is the inclination, $\omega$ is the argument of periastron, $\Omega$ is the longitude of nodes, $M_0$ is the mean anomaly at epoch $\tau_0$ = 2014.0896 (MJD\footnote{Modified Julian Date = Julian Date - 2400000.5} = 56675.982), $q \equiv {\rm M}_2/{\rm M}_1$ is the system mass ratio, $V_{COM}$ is the center of mass (systemic) radial velocity in {\kms}, and $d$ is the distance in pc.  The primary radial velocity as a function of time $t$, $V_1(t)$, is
\begin{equation}
V_1(t) = K_1\left[e\cos{\omega} + \cos{(T(t)+\omega)}\right]+ V_{COM}
\end{equation}
where
\begin{equation}
K_1 = \frac{2\pi{a}\sin{i}}{P\sqrt{1-e^2}} \frac{q}{1+q}
\end{equation}
and the true anomaly $T(t)$ is related to the eccentric anomaly $E(t)$ through
\begin{equation}
\tan{\frac{T(t)}{2}} = \sqrt{\frac{1+e}{1-e}}\tan{\frac{E(t)}{2}}
\end{equation}
which is solved by Kepler's Equation:
\begin{equation}
M(t) - M_0 = 2\pi\frac{t-{\tau}_0}{P} = E(t) - e\sin{E(t)}.
\end{equation}
The angular separation vector from primary component to secondary component,  $\vec{\rho}$ = ($\Delta\alpha$(t), $\Delta\delta$(t)) is determined from
\begin{align}
\Delta\alpha(t) = \frac{a}{d} \left[A(\cos{E(t)} - e) + F\sqrt{1-e^2}\sin{E(t)}\right] \\
\Delta\delta(t) = \frac{a}{d} \left[B(\cos{E(t)} - e) + G\sqrt{1-e^2}\sin{E(t)}\right]
\end{align}
where $\Delta\alpha$ and $\Delta\delta$ are the angular separations on sky measured in arcseconds, and $A$, $B$, $F$ and $G$ are the Thiele-Innes constants \citep{1907Obs....30..310I,1927BAN.....3..261V}:
\begin{align}
A & = \cos{\omega}\cos{\Omega} - \sin{\omega}\sin{\Omega}\cos{i} \\
B & = \cos{\omega}\sin{\Omega} + \sin{\omega}\cos{\Omega}\cos{i} \\
F & = -\sin{\omega}\cos{\Omega} - \cos{\omega}\sin{\Omega}\cos{i} \\
G & = - \sin{\omega}\sin{\Omega} + \cos{\omega}\cos{\Omega}\cos{i}.
\end{align}
Note that the total system mass (M$_{tot} = a^3/P^2$ in solar masses) and component masses (M$_1$ = M$_{tot}$/[1$-q$], M$_2$ = $q$M$_{1}$) {could in principle} be uniquely inferred from these parameters if sufficiently constrained.

We selected an initial parameter set that visually coincided with the observations through manual experimentation. 
We then computed a chain of 10$^7$ parameter sets {using the Metropolis-Hastings algorithm \citep{1953JChPh..21.1087M,HASTINGS01041970}}, at each step varying parameter $\theta_j \rightarrow \theta_j^{\prime}$ by drawing a random offset from a normal distribution
\begin{equation}
P(\theta_j^{\prime}|\theta_j) \propto {\rm e}^{-\frac{(\theta_j^{\prime}-\theta_j)^2}{2\beta{j}^2}}
\end{equation}
where $\vec{\beta}$ is the set of jump steps.\footnote{We used in initial set $\vec{\beta}$ = (3~yr, 0.5~AU, 0.2,10$\degr$,10$\degr$,10$\degr$,10$\degr$, 0.2, 1.0~{\kms}, 1.0~pc).} We applied additional parameter constraints of 0.5~yr $< P <$ 30~yr, $e < {0.8}$ and {4}~pc $< d <$ {8}~pc to eliminate improbable regions of parameter space; {note that the eccentricity cutoff is beyond the $e \approx 0.6$ limit suggested in empirical data by \citet{2011ApJ...733..122D}}.  We also {limited the component masses} to 0.055~{\msun} $<$ M$_1$ $<$ 0.15~{\msun} given the spectral classification of the primary and lack of Li~I absorption in its optical spectrum (B15; I15), and M$_{2} < 0.075$~{\msun} given the substellar nature of the secondary.  
{Orbit models} were compared {to the data using a $\chi^2$ statistic that combined both RV and relative astrometric measurements:}
\begin{equation}
\begin{aligned}
\chi^2 = \sum_{i=1}^{N_{RV}}\frac{(RV_i^{(obs)}-RV_i^{(model)})^2}{\sigma_{RV,i}^2} + \\
\sum_{j=1}^{N_{ast}}\frac{(\Delta\alpha_j^{(obs)}-\Delta\alpha_j^{(model)})^2}{\sigma_{\Delta\alpha,j}^2} + \\
\sum_{j=1}^{N_{ast}}\frac{(\Delta\delta_j^{(obs)}-\Delta\delta_j^{(model)})^2}{\sigma_{\Delta\delta,j}^2} 
\end{aligned}
\end{equation}
{where $N_{RV}$ and $N_{ast}$ are the number of RV and astrometric measurements, respectively, and $\sigma$ the measurement errors. Model values were calculated at the same epochs as the observations.  The criterion to adopt successive parameter sets was $U(0,1) \leq e^{-0.5(\chi^2_{(j^{\prime})}-\chi^2_{(j)})}$, where $U(0,1)$ is a random number drawn from a uniform distribution between 0 and 1. 
We compared models to data
after each individual parameter change rather than changing all parameters, a procedure we found greatly improved the acceptance rate, which declined from 10\% to 1\% through the chain. 
Separate fits were made to the NIRSPEC + NIRC2 and Hamilton +NIRC2 measurements, and to all of the data. Note that in the last case we did not take into account a possible velocity offset between the RV datasets \citep{2005AJ....129.1706F}.}

{To test for convergence, we monitored parameter autocorrelation and the evolution of subchain variance through the chain. We also generated $M$ = 10 chains of length $2n$ = 10$^6$ steps on the Hamilton + NIRC2 data, varying the initial parameter set $\vec{\theta}_{init}$ as 
\begin{equation}
\vec{\theta}_{init} = \vec{\theta}_{p} + N(0,1)\vec{\sigma}_p
\end{equation}
where $\vec{\theta}_{p}$ and $\vec{\sigma}_p$ are the median and half quantile ranges of the posterior parameter distributions (see below) and $N(0,1)$ is a normal distribution centered on zero with unit variance. We quantified the within chain and between chain variance of all parameters using the \citet{doi:10.1214/ss/1177011136} scale reduction factor
\begin{equation}
R^2_k = 1 + \frac{1}{n}\left(\frac{B_k}{W_k}-1\right)
\end{equation}
where $W_k = \bar{\xi}_k$ is the within chain variance for parameter $\theta_k$, equal to the average of parameter variances $\xi_{jk} = \frac{1}{n-1}\sum_{i=1}^n(\theta_{ijk}-\bar{\theta}_{jk})^2$ for each of the ten chains using the second half of the chains as the sample; and  
$B_k = \frac{n}{M-1}\sum_{j=1}^{M}(\bar{\theta}_{jk} - \bar{\bar{\theta}}_k)^2$ is the between chain variance, equal to the variance in parameter averages $\bar{\theta}_{jk}$. 
Scale reduction factors were within 5\% of unity for all modeled parameters with the exception of $P$ (1.35), $a$ (1.37) and $\omega$ (1.51).  As described below, these three parameters were weakly constrained, and this analysis suggests that the primary MCMC chains may not converge for our limited datasets, re-emphasizing that the orbital results presented here should be considered preliminary.
}

\subsection{Results}

\begin{figure*}
\epsscale{1.1}
\centering
\includegraphics[width=10cm]{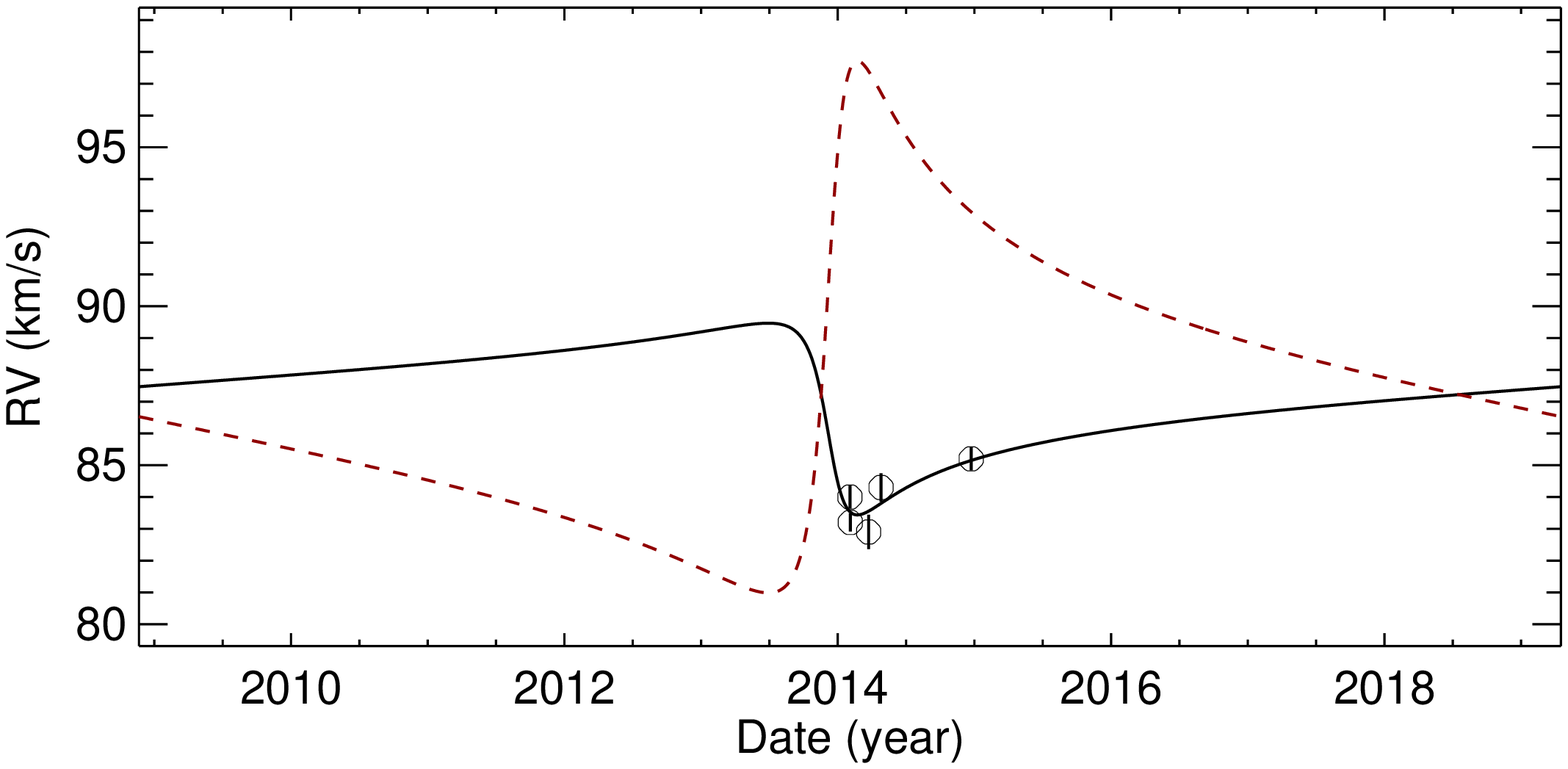}
\includegraphics[width=5cm]{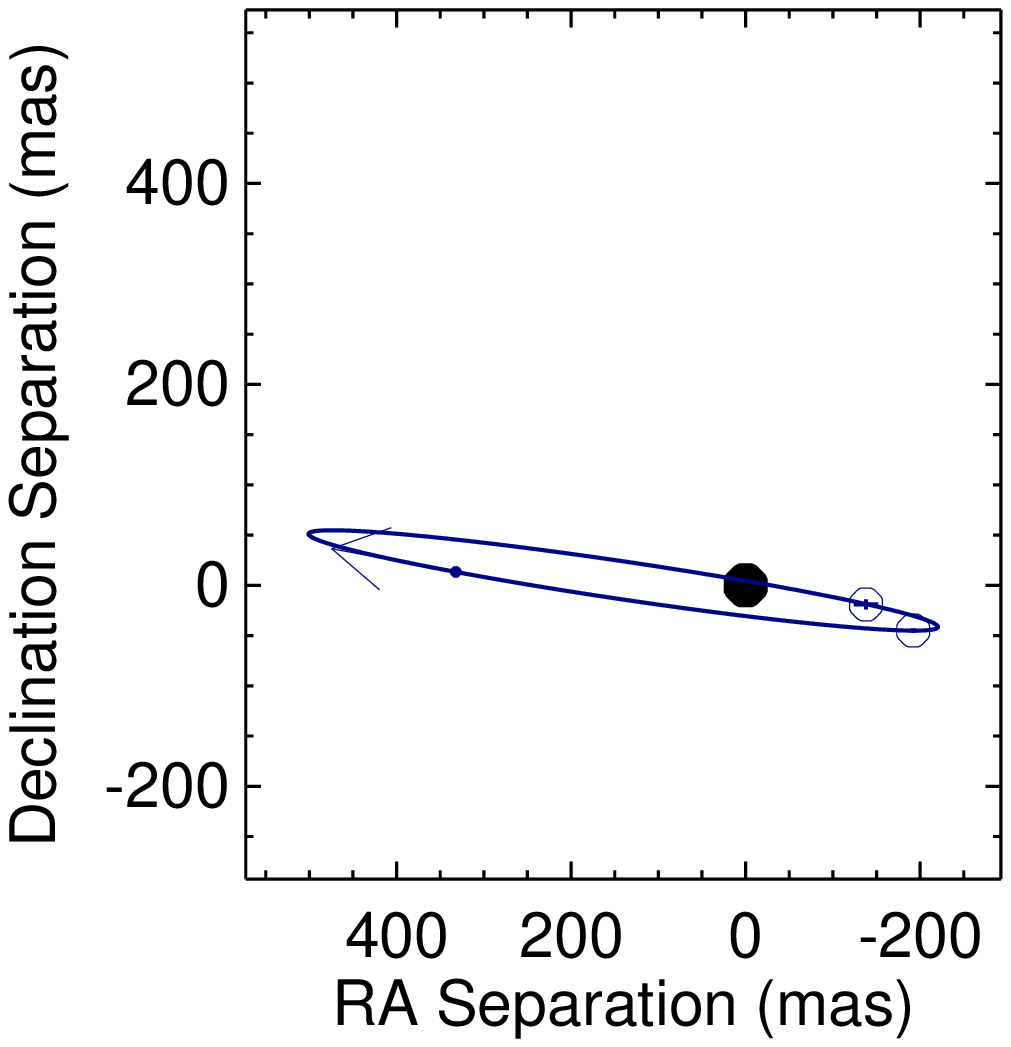} \\
\includegraphics[width=10cm]{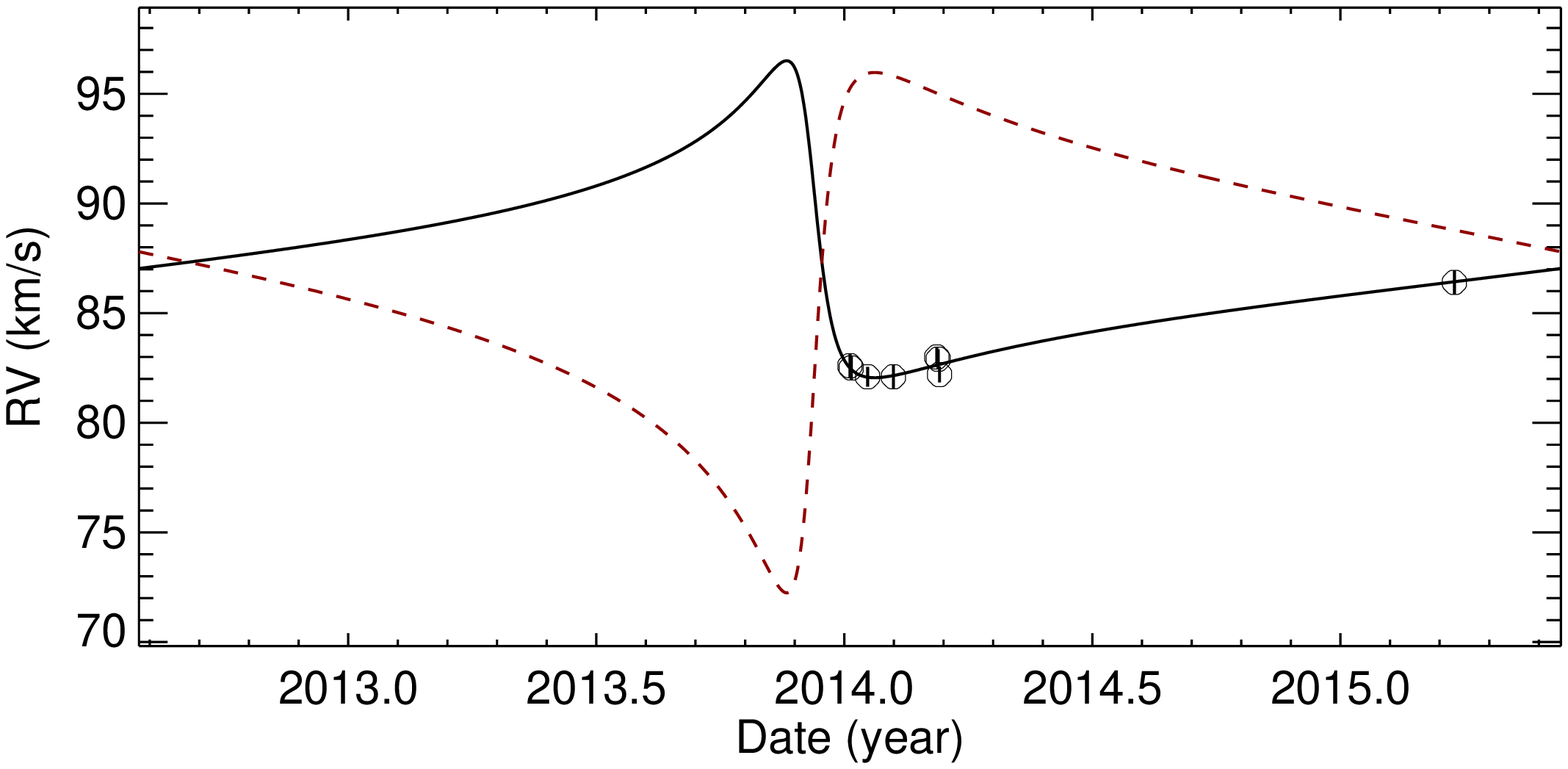}
\includegraphics[width=5cm]{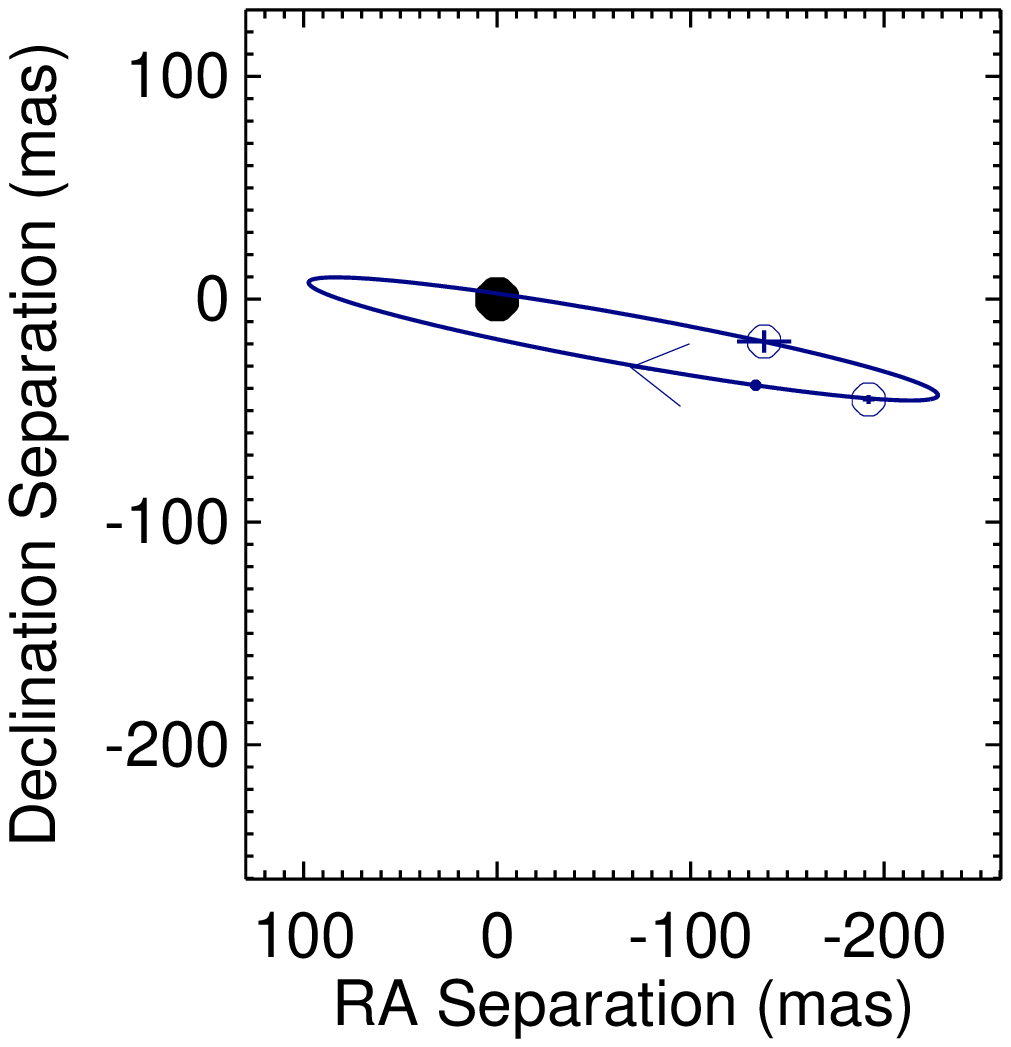} \\
\includegraphics[width=10cm]{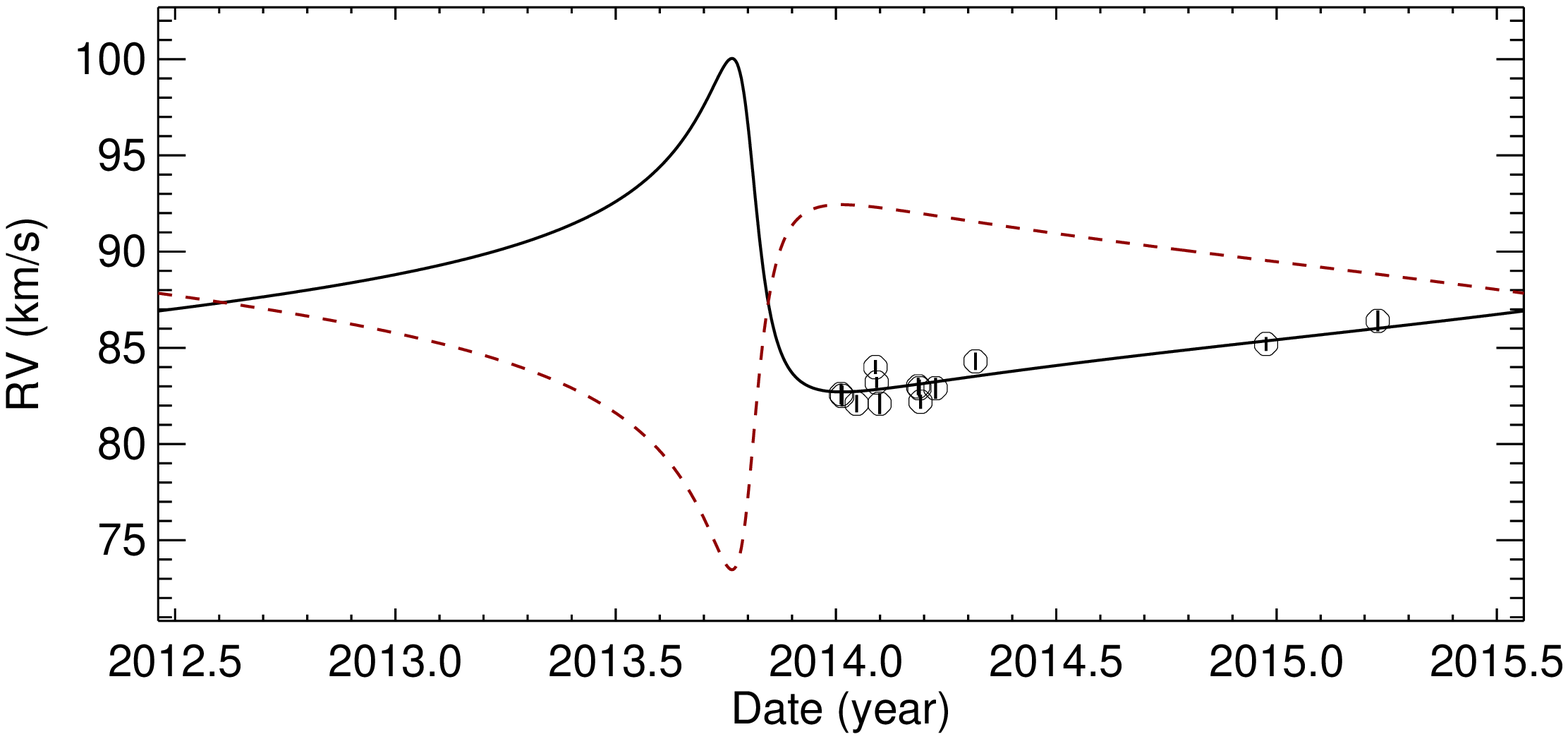}
\includegraphics[width=5cm]{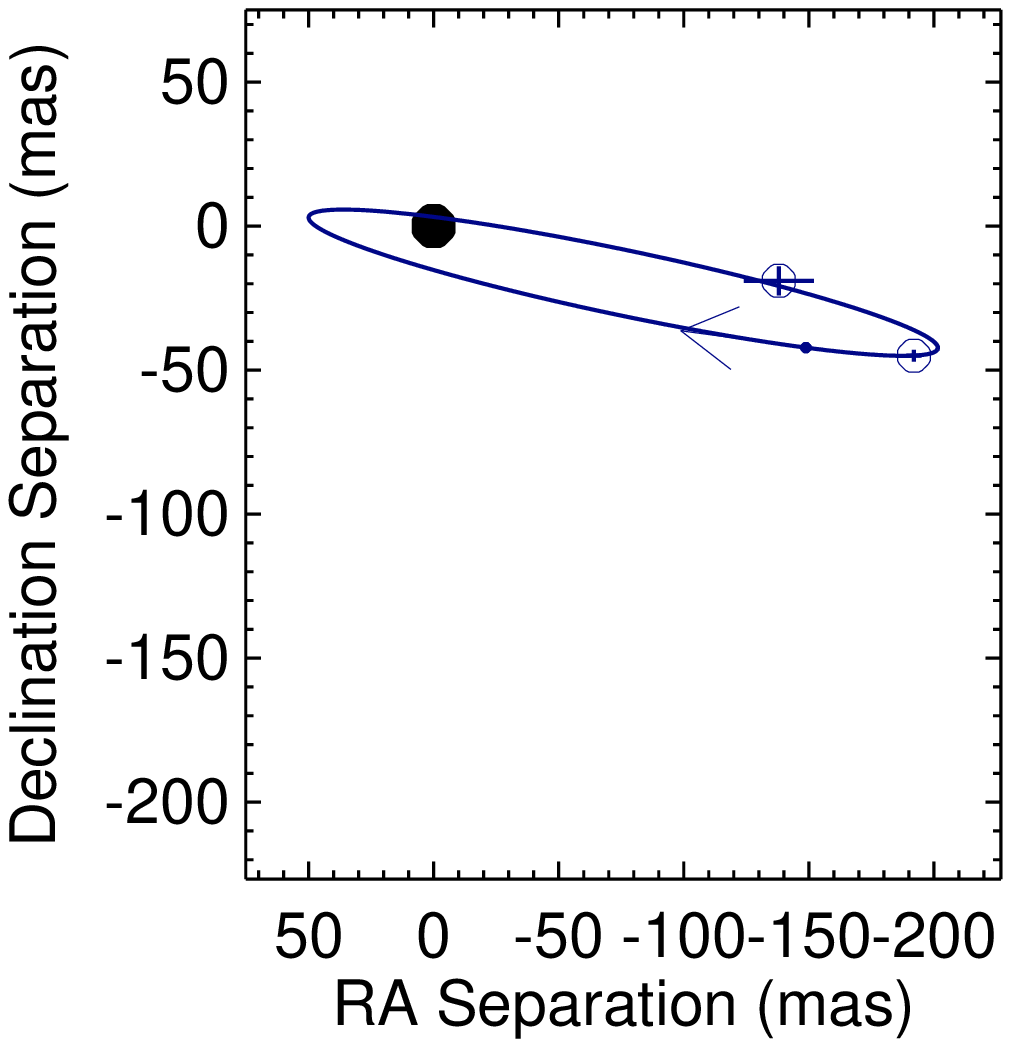} \\
\caption{Best-fit orbit from MCMC analysis based on NIRSPEC + NIRC2 data (top), Hamilton Spectrograph + NIRC2 data (middle), {and all data (bottom)}. The left panels show the radial motion of both the primary (solid black line) and secondary (red dashed line) motion compared to primary RV measurements (open circles with error bars). The right panels show the orbital motion of the secondary (blue line) relative to the primary (black dot at the origin) projected on the sky, compared to NIRC2 measurements (open circles; error bars are small on this scale). The arrow indicates the direction of orbital motion at apoapse. 
{Note the different scales for the different fits.} 
\label{fig:orbit}}
\end{figure*}

\begin{figure}
\epsscale{1.0}
\centering
\plotone{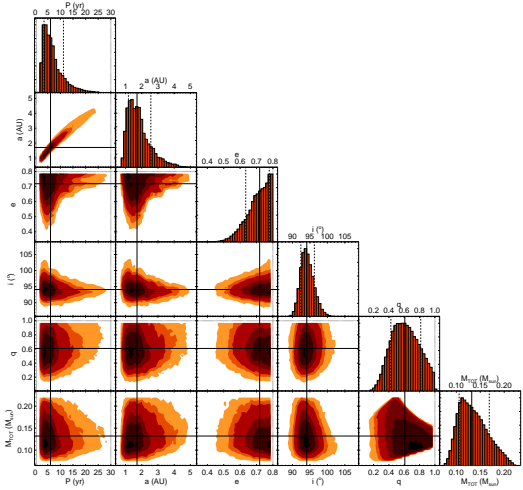}
\caption{Parameter distributions and correlations (triangle plot) for period ($P$), semi-major axis ($a$), eccentricity ($e$), 
inclination ($i$), mass ratio ($q$) and total system mass ($M_{tot}$) based on our MCMC orbital analysis for the NIRSPEC + NIRC2 data.
The fits assume a weak constraint on eccentricity (0 $\leq e \leq$ 0.8) and period (0.5~yr $\leq P \leq$ 30~yr).
Contour plots show two-dimensional $\chi^2$ distributions for parameter pairs, highlighting correlations.
Normalized histograms at the ends of rows are marginalized over all other parameters.
Median values are indicated by solid lines in all panels; 16\% and 84\% quantiles are indicated by dashed lines in the histograms.  Imposed parameter limits for $P$ and $e$ are indicated by dotted lines.
\label{fig:orbit_parameters}}
\end{figure}

\begin{figure}
\epsscale{1.0}
\centering
\plotone{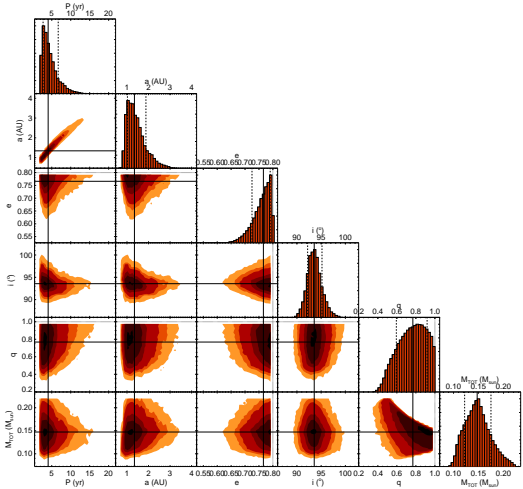}
\caption{Same as Figure~\ref{fig:orbit_parameters} for the Hamilton Spectrograph + NIRC2 data.
\label{fig:orbit_parameters_hamilton}}
\end{figure}

\begin{figure}
\epsscale{1.0}
\centering
\plotone{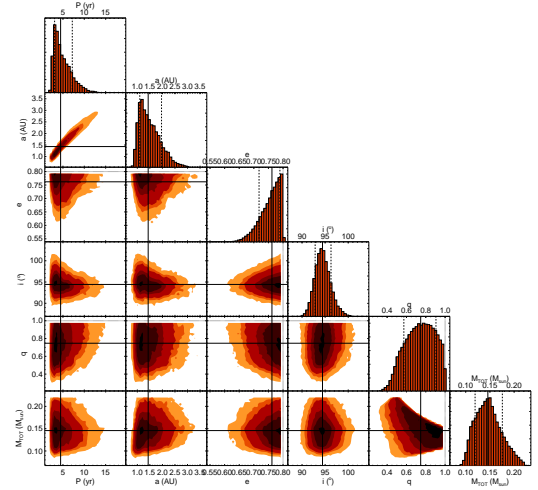}
\caption{Same as Figure~\ref{fig:orbit_parameters} for all RV and imaging data.
\label{fig:orbit_parameters_all}}
\end{figure}

Figure~\ref{fig:orbit} shows the best-fit orbits from our separate analyses of the RV and imaging datasets.
{Table~\ref{tab:orbit_parameters} lists the best-fit orbital and component parameters, as well as the median and 16\% and 84\% quantiles of the parameter distributions, after eliminating the first 10\% of each chain. 
Figures~\ref{fig:orbit_parameters} through~\ref{fig:orbit_parameters_all} display the distributions and correlations of parameters $P$, $a$, $e$, $i$, $q$ and M$_{tot}$ for all three datasets.}

\begin{deluxetable*}{lcccccc}
\tablecaption{Parameters from Orbital Analysis \label{tab:orbit_parameters}}
\tabletypesize{\small}
\tablewidth{0pt}
\tablehead{
 & \multicolumn{2}{c}{NIRSPEC+NIRC2} & \multicolumn{2}{c}{Hamilton+NIRC2} & \multicolumn{2}{c}{All Data} \\
 \cline{2-3} \cline{4-5} \cline{6-7}
\colhead{Parameter} &
\colhead{Best-fit} &
\colhead{Median} &
\colhead{Best-fit} &
\colhead{Median} &
\colhead{Best-fit} &
\colhead{Median} }
\startdata
\hline
\multicolumn{7}{c}{Modeled Parameters} \\
\hline
Best $\chi^2$  & 5.83 &  \nodata  & 2.84 &  \nodata  & 27.6  & \nodata \\
$P$\tablenotemark{a} (yr) & 10.4 &  6.1$^{+5.1}_{-2.6}$ & 2.9 & 4.1$^{+2.7}_{-1.3}$ & 3.1 & 4.5$^{+2.7}_{-1.4}$ \\
$a$ (AU) & 2.4  & 1.7$^{+0.9}_{-0.5}$ & 1.1 &  1.3$^{+0.5}_{-0.3}$ & 1.1 &  1.4$^{+0.5}_{-0.3}$ \\
$e$\tablenotemark{a} & 0.79 &  0.72$^{+0.06}_{-0.09}$ & 0.79 &  0.77$^{+0.02}_{-0.04}$ & 0.80 &  0.76$^{+0.03}_{-0.04}$ \\
$i$ ($\degr$) & 92.1 & 94.1$^{+2.2}_{-1.7}$ & 92.8 &  93.6$^{+1.6}_{-1.4}$ & 94.9 &  94.4$^{+1.8}_{-1.6}$ \\
$\omega$ ($\degr$) & 109 &  83$^{+20}_{-24}$ & 70 &  76$^{+17}_{-17}$ & 54 &  74$^{+16}_{-16}$ \\
$\Omega$ ($\degr$) & 81.7 & 82.2$^{+2.3}_{-2.1}$ & 82.4 &  82.7$^{+2.1}_{-2.1}$ & 82.2 &  83.3$^{+2.2}_{-2.2}$ \\
$M_0$ ($\degr$) & 5 & 20$^{+15}_{-11}$ & 10 &  13$^{+8}_{-6}$ & 24 &  17$^{+8}_{-7}$ \\
$q$  & 0.36 &  0.61$^{+0.21}_{-0.18}$ & 0.61 &  0.77$^{+0.15}_{-0.17}$ & 0.91 &  0.75$^{+0.16}_{-0.17}$ \\
$V_{COM}$ ({\kms})  & 87.2 &  87.3$^{+0.9}_{-0.9}$ & 87.3 &  87.4$^{+0.8}_{-0.8}$ & 87.4 &  87.5$^{+0.7}_{-0.7}$ \\
$d$\tablenotemark{a} (pc)  & 4.5 & 6.7$^{+0.8}_{-1.0}$  & 4.5 &  5.4$^{+0.8}_{-0.7}$  & 6.8 &  6.2$^{+0.7}_{-0.7}$ \\
\hline
\multicolumn{7}{c}{Inferred Parameters} \\
\hline
$M_{tot}$ ({\msun}) & 0.13 &  0.13$^{+0.04}_{-0.03}$  & 0.17 &  0.15$^{+0.03}_{-0.03}$  & 0.16 &  0.15$^{+0.03}_{-0.03}$ \\
$M_{1}$ ({\msun}) & 0.098 &  0.080$^{+0.032}_{-0.018}$  & 0.105 &  0.082$^{+0.026}_{-0.015}$  & 0.081 &  0.081$^{+0.028}_{-0.016}$ \\
$M_{2}$ ({\msun}) & 0.035 &  0.051$^{+0.013}_{-0.014}$  & 0.064 &  0.064$^{+0.008}_{-0.010}$  & 0.074 &  0.062$^{+0.009}_{-0.011}$ \\
$K_1$ ({\kms}) & 3.0 &  4.4$^{+1.2}_{-1.0}$  & 7.2 &  6.3$^{+1.1}_{-1.0}$ & 8.7 &  6.0$^{+1.1}_{-0.8}$ \\
$K_2$ ({\kms}) & 8.4  &  7.3$^{+2.2}_{-1.6}$  & 11.9  &  8.5$^{+2.1}_{-1.6}$ & 9.5  &  8.2$^{+2.1}_{-1.6}$ 
\enddata
\tablenotetext{a}{Parameter was constrained to a limited value range in MCMC analysis.}
\end{deluxetable*}

The best-fit orbits are exceptionally good fits for the individual RV datasets, with $\chi^2$ = {5.83} for nine data points {(zero degrees of freedom)} for NIRSPEC + NIRC2 and $\chi^2$ = {2.84} for twelve data points {(three degrees of freedom)} for Hamilton + NIRC2. 
Note that these low $\chi^2$ values largely reflect the underconstained nature of the solution in orbital phase; 
{i.e., a relatively large family of solutions is formally consistent with these data. 
The combined dataset yields a somewhat poorer best fit, with modest disagreement between contemporary NIRSPEC and Hamilton RVs driving the $\chi^2$ values.  Nevertheless, this solution is nearly identical to that of the Hamilton + NIRC2 dataset.
While
the best-fit solutions between the NIRSPEC + NIRC2 and the Hamilton + NIRC2 datasets are distinct, the median parameters are consistent within the uncertainties for all datasets. Since parameters inferred from the Hamilton + NIRC2 data are best constrained, we refer to these in the following discussion.} 

Several of the parameters are very well constrained, {most notably} the orbital inclination {($i$ = 93$\fdg$6$^{+1\fdg6}_{-1\fdg4}$)}, 
the longitude of ascending node {($\Omega$ = 82$\fdg$7$\pm$2$\fdg$1)} and 
the center of mass radial velocity {($V_{COM}$ = +87.4$\pm$0.8~{\kms})}.
{The tight constraints on $i$ and $\Omega$ stem from the near-radial motion of the secondary between the two NIRC2 epochs. The position angle of the secondary changed by only 5$\degr$$\pm$2$\degr$ between these two observations but the secondary has moved 40\% further away; this is possible only when the orbit is viewed very close to edge-on or $e$ $\approx$ 1. As discussed below, even with an eccentricity at the proscribed limits, the inclination remains close to 90$\degr$.
Similarly, $\Omega$ is constrained by the orientation of the orbit on the sky.
The tight constraint on $V_{COM}$ arises from the small variance in observed RVs and consistency in the values measured in the earlier epochs.  The NIRSPEC + NIRC2 and Hamilton + NIRC2 datasets yield nearly identical values for all three of these parameters.} 

\begin{figure}
\epsscale{1.1}
\centering
\includegraphics[width=8cm]{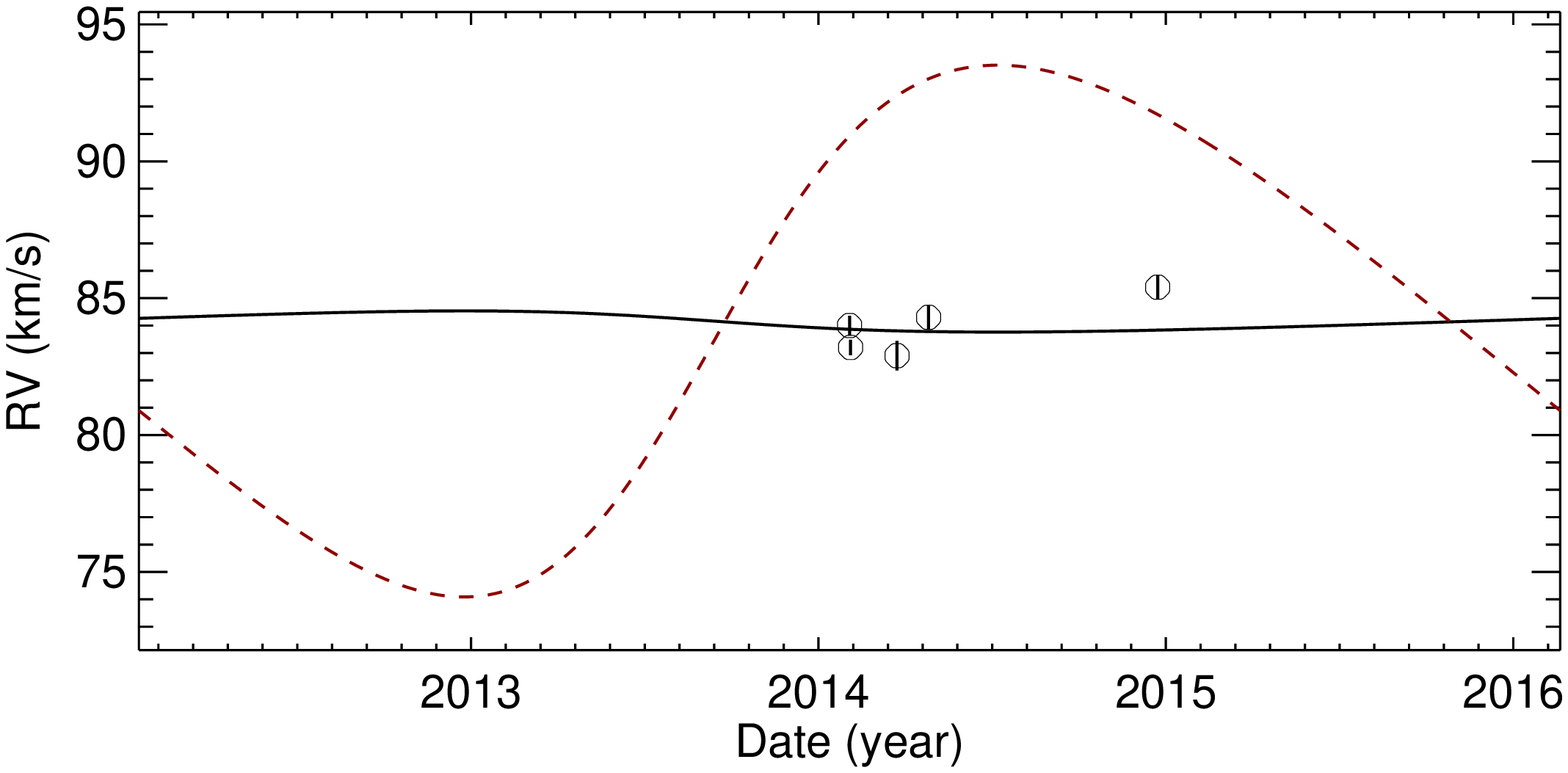}
\includegraphics[width=8cm]{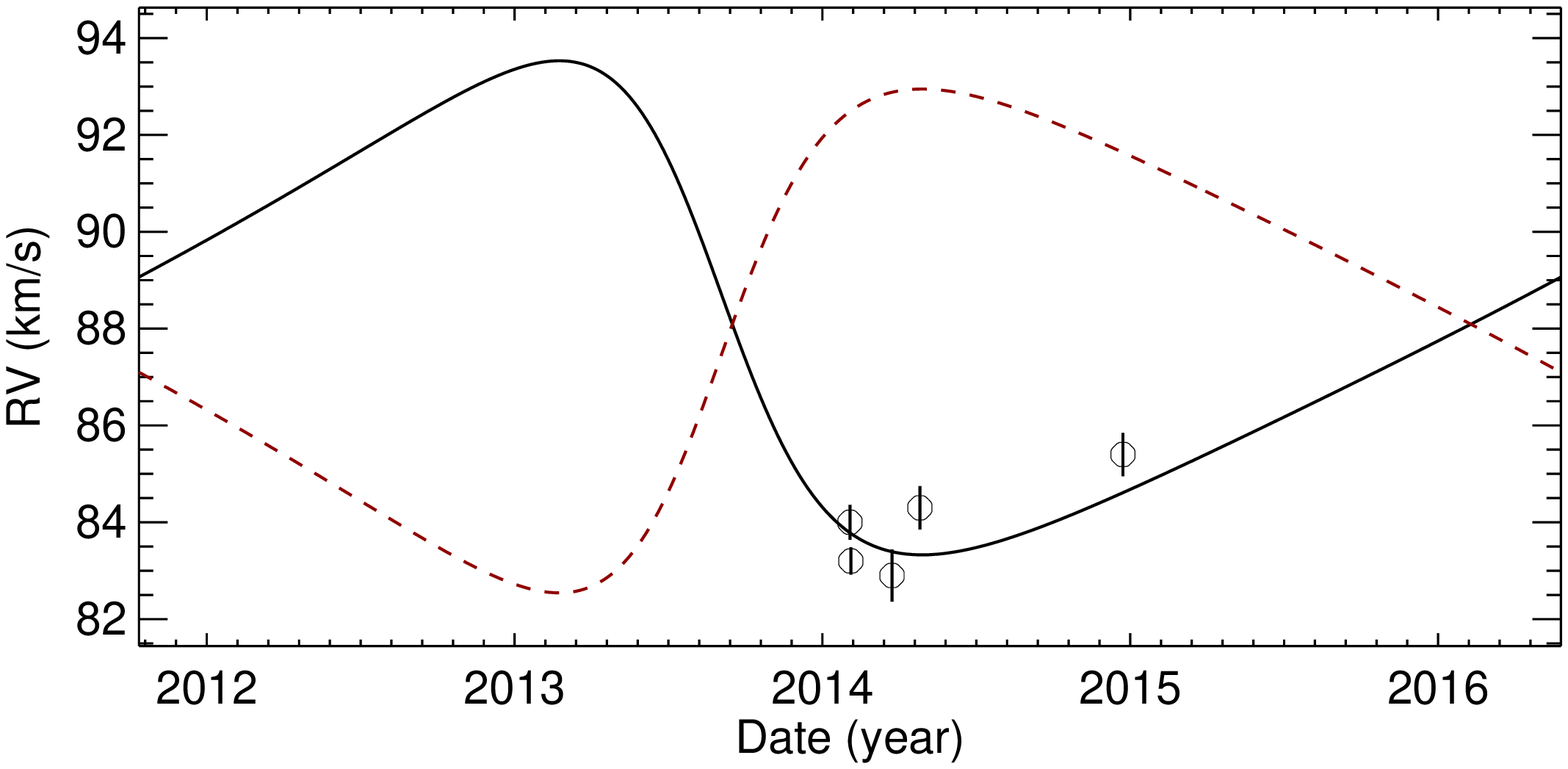} \\
\includegraphics[width=8cm]{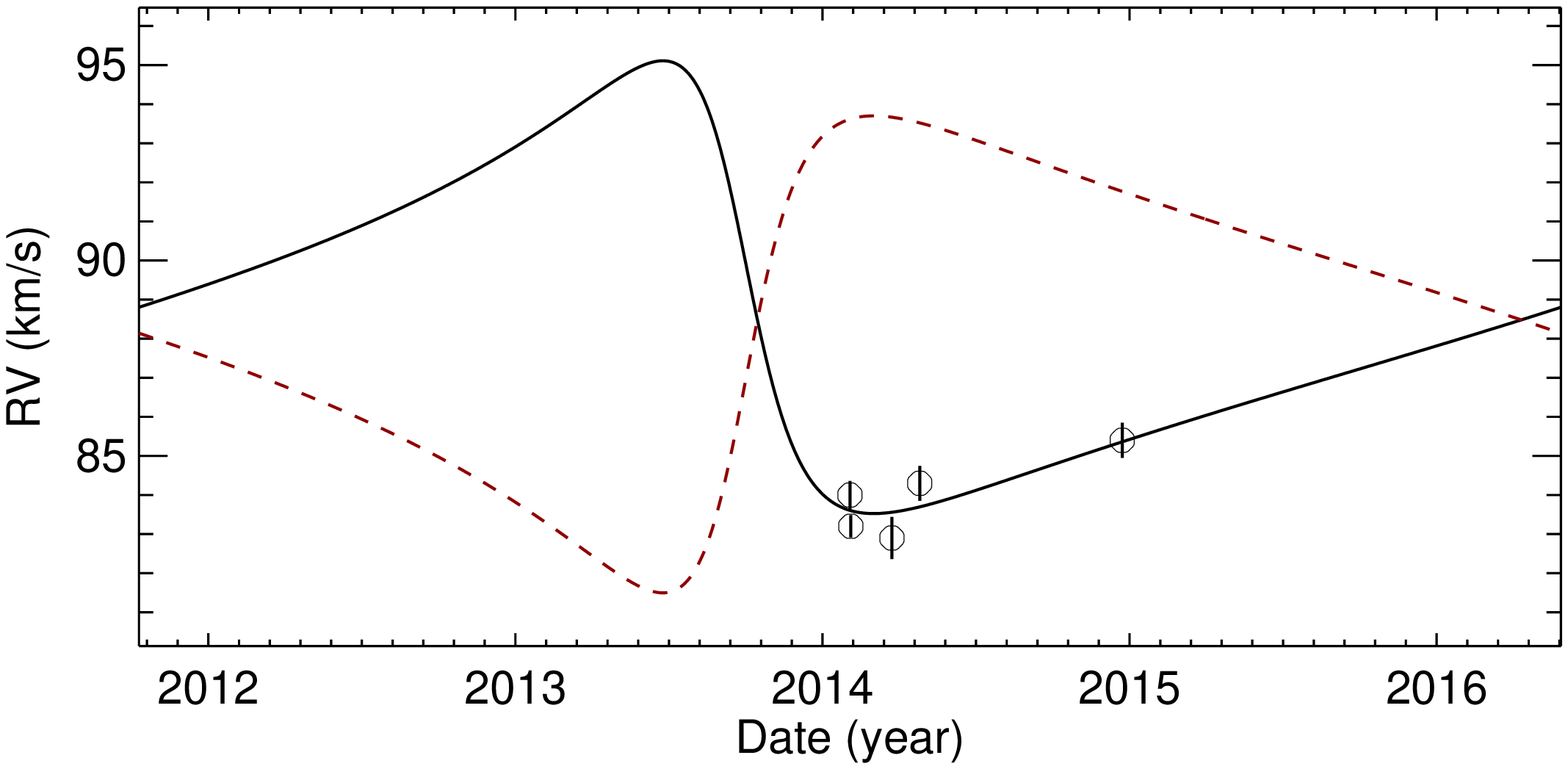}
\includegraphics[width=8cm]{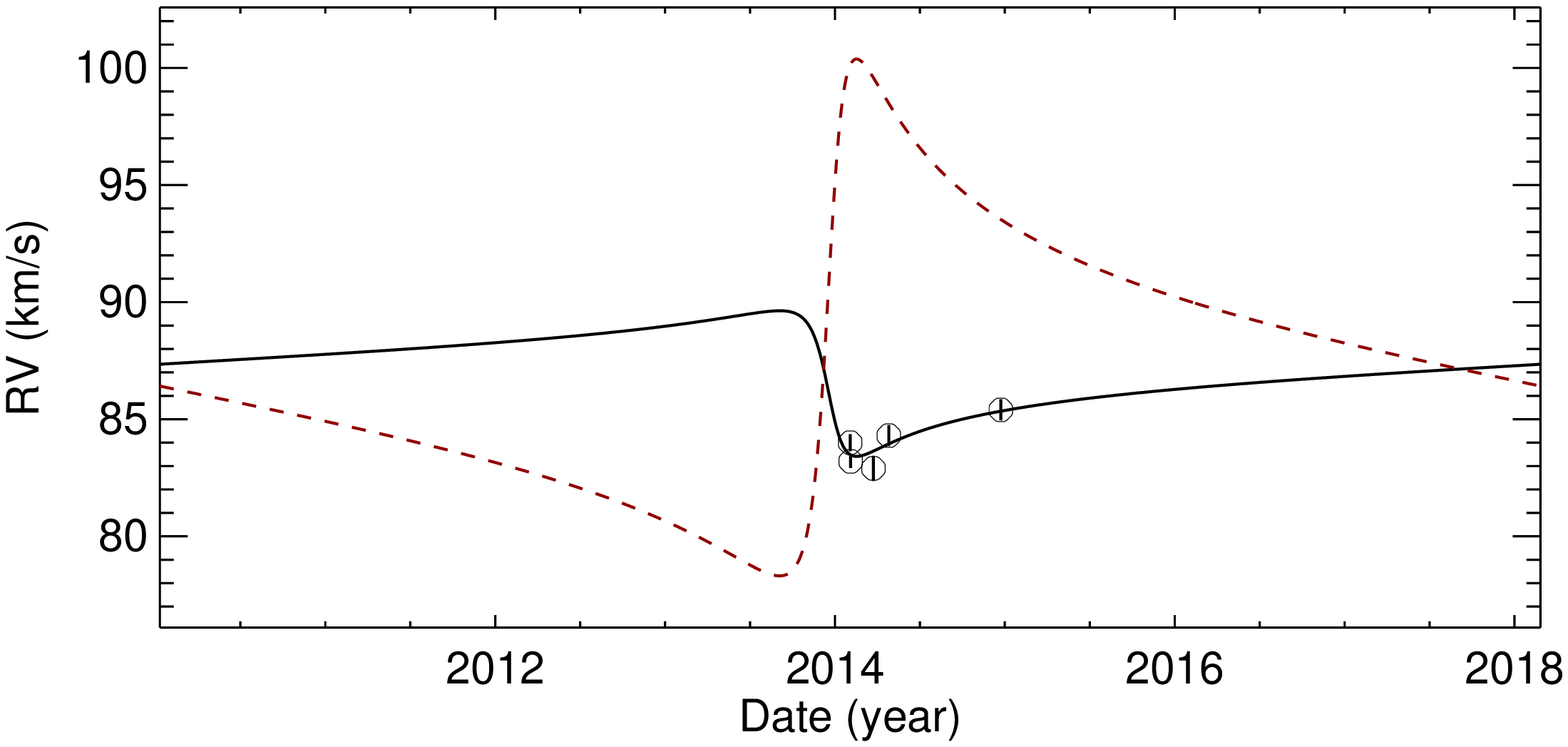}
\caption{Predicted primary (black solid line) and secondary (red dashed line) radial velocities based on best-fit orbits with eccentricity fixed at (from top to bottom): 0.2, 0.4, 0.6 and 0.8, compared to NIRSPEC data. Compare to Figure~\ref{fig:orbit}.  
\label{fig:orbit_eccentricity}}
\end{figure}

{Parameter distributions for $P$, $e$ and $q$ are more weakly constrained, and the latter two abut the imposed limits, emphasizing that we do not yet have sufficient coverage of the orbit of {\namesh}AB to robustly determine them. To gain some insight on how the quality of the fits vary as $e$ changes, we performed additional MCMC
chains on Hamilton + NIRC2 data using models with  fixed $e$ = 0.2, 0.4, 0.6 and 0.8.  Figure~\ref{fig:orbit_eccentricity} and Table~\ref{tab:orbit_parameters_eccentricity} display the results of this experiment. 
We found that all parameters remain consistent with the unconstrained MCMC analysis, although $q$ and M$_2$ values are modestly smaller for the $e$ = 0.2.
Parameter uncertainties are considerably larger for the $e$ = 0.8 case, in part reflecting the lower $\chi^2$ values of viable solutions.
It is clear that the $e$ = 0.2 and $e$ = 0.4 models are poor representations of the RV data, with best-fit $\chi^2$ values that are significantly worse and can be eliminated with $>$95\% confidence as compared to the unconstrained model.\footnote{This confidence level was computed using the F-test probability distribution function, comparing the unconstrained best-fit model ($\chi^2$ = 2.84, 3 degrees of freedom) to the constrained models ($\chi^2$ = 39.3 and 25.1, 4 degrees of freedom).}  
This analysis supports an eccentic orbit for {\namesh}AB, but a firm constraint on its value should emerge with further monitoring.}

\begin{deluxetable*}{lcccc}
\tablecaption{Parameters from Orbital Analysis of Hamilton + NIRC2 Data with Eccentricity Fixed \label{tab:orbit_parameters_eccentricity}}
\tabletypesize{\small}
\tablewidth{0pt}
\tablehead{
\colhead{Parameter} &
\colhead{$e$ = 0.2} &
 \colhead{$e$ = 0.4} &
 \colhead{$e$ = 0.6} &
 \colhead{$e$ = 0.8} 
 }
\startdata
Best $\chi^2$  & 39.3 & 25.1 & 10.4 & 2.8 \\
$P$\tablenotemark{a} (yr) & 4.4$^{+0.8}_{-0.7}$ & 4.3$^{+1.0}_{-0.8}$ & 4.2$^{+1.2}_{-0.9}$ &  3.8$^{+3.0}_{-1.3}$ \\
$a$ (AU) &  1.40$^{+0.18}_{-0.16}$ &  1.38$^{+0.23}_{-0.18}$ &  1.4$^{+0.3}_{-0.2}$ & 1.3$^{+0.6}_{-0.3}$ \\
$i$ ($\degr$) &  94.1$^{+1.5}_{-1.3}$ & 94.4$^{+1.4}_{-1.3}$ &  94.2$^{+1.5}_{-1.3}$ &  93.5$^{+2.0}_{-1.6}$ \\
$q$\tablenotemark{a}  & 0.69$^{+0.19}_{-0.20}$ &  0.80$^{+0.14}_{-0.18}$ &  0.82$^{+0.13}_{-0.17}$ &  0.73$^{+0.16}_{-0.18}$ \\
$V_{COM}$ ({\kms})  &   85.9$^{+0.5}_{-0.6}$ &  86.5$^{+0.5}_{-0.5}$ &  87.0$^{+0.5}_{-0.5}$ & 87.3$^{+0.9}_{-0.9}$ \\
$M_{tot}$ ({\msun}) &   0.14$^{+0.03}_{-0.03}$  &  0.14$^{+0.03}_{-0.02}$  &  0.15$^{+0.02}_{-0.02}$  &  0.15$^{+0.03}_{-0.03}$  \\
$M_{1}$ ({\msun}) &   0.083$^{+0.032}_{-0.018}$  &  0.079$^{+0.024}_{-0.014}$  &  0.079$^{+0.022}_{-0.013}$  &   0.083$^{+0.028}_{-0.017}$  \\
$M_{2}$ ({\msun}) &   0.059$^{+0.011}_{-0.014}$  &   0.064$^{+0.008}_{-0.011}$  &  0.066$^{+0.007}_{-0.010}$  &  0.062$^{+0.009}_{-0.011}$  
\enddata
\tablenotetext{a}{Parameter was constrained to a limited value range in MCMC analysis.}
\end{deluxetable*}

\section{Discussion}

Our follow-up observations of {\namesh}AB verify both the magnetic and binary nature of this nearby system, and provide first constraints on the physical properties of the components.  While these first measurements of orbital motion {do not} robustly constrain system or component masses, they do constrain the orbital geometry, most notably inclination, which is within a few degrees of edge-on.  The scale of this system is such that it is unlikely to eclipse; nevertheless, the orbit orientation allows us to make some constraints on the rotational properties of the primary. 
The orbital and primary rotational axes of main sequence binaries with close separations ($\lesssim$20\,AU) are generally aligned  \citep{1974ApJ...190..331W,1994AJ....107..306H}. The only very low mass binary for which spin-orbit alignment has been tested, 2MASSW J0746425+200032AB, shows alignment between orbital and rotational axes of both components to within 5$\degr$ \citep{2013A&A...554A.113H}. Assuming similar alignment of the rotational and orbital axes of {\namesh}A and a radius of 0.1\,R$_{\odot}$, our {\vsini} measurement implies a rotation period of {14.2$\pm$0.7~hr}. This {value} is remarkably close to the marginally-indicated variability period from white light monitoring reported in B15, 14.00$\pm$0.05~hr for a 1.3$\pm$0.5\% variability amplitude. {The rotation period is} somewhat slower than the mean for comparably classified sources \citep{2011ApJ...727...56I} and considerably slower than most highly variable L and T dwarfs \citep{2014ApJ...793...75R,2015ApJ...799..154M}. Its slow rotation rate and orientation may explain the weaker than average magnetic emission of {\namesh}A, in both optical and radio bands, and its somewhat low flaring rate compared to other late M dwarfs.  In addition, with a 1--2~AU semimajor axis, magnetospheric interaction between primary and secondary is unlikely to play a role in driving magnetic emission in either source \citep{2009ApJ...699L.148S}.

A more promising role for {\namesh}AB is as a testbed for brown dwarf evolutionary models. As demonstrated by several studies (e.g., \citealt{2008ApJ...689..436L,2010ApJ...722..311L,2010ApJ...711.1087K,2014ApJ...790..133D}), the inferred masses and atmospheric parameters of several very low mass binaries diverge to varying degrees from evolutionary model predictions. Limiting factors for such analyses include lack of knowlege of detailed spectra for individual components, inability to measure component masses,  low quality distance determinations, and/or lack of independent age determinations. While the last factor may be challenging to overcome for {\namesh}AB, the proximity and detection of reflex motion in both astrometric and radial coordinates allows the first three to be addressed, and we anticipate that this system will provide a high-quality test of models in the next 5-10 years.

Finally, we note that despite being able to detect reflex motion in the primary, the inferred close passage of {\namesh}AB to the Sun deduced by \citet{2015ApJ...800L..17M} on the basis of the system kinematics is not ruled out.  The systemic motion estimated here is about 4\,{\kms} larger than that assumed by these authors, so the impact parameter is roughly 5\% smaller. The astrometric perturbation of the primary induced by the secondary, of order 100~mas, is sufficiently slow to play little role in modulating the parallax or proper motion of the system significantly given current measurement uncertainties.  Improved monitoring of the systemic orbital motion will be needed to make a more precise estimate of the geometry and timescale of {\namesh}AB's closest approach, and the corresponding perturbation it has made on our Oort cloud comet population.

\acknowledgements
The authors thank  Randy Campbell, Heather Hershley, and Marc Kassis at Keck Observatory; and
Pavl Zachary and Shawn Stone at Lick Observatory for their assistance with the observations.
{We thank our anonymous referee for her/his/their helpful comments, particularly on the MCMC analysis.}
C.M. acknowledges funding support from the National Science Foundation under award No.\ AST-1313428. 
The material is based upon work supported by the National Aeronautics and Space Administration under Grant No. NNX15AI75G.
This research has made use of the SIMBAD database,
operated at CDS, Strasbourg, France;
NASA's Astrophysics Data System Bibliographic Services;
the M, L, T, and Y dwarf compendium housed at DwarfArchives.org;
and the SpeX Prism Libraries at \url{http://www.browndwarfs.org/spexprism}.
Research at Lick Observatory is partially supported by a generous gift from Google.
The authors recognize and acknowledge the very significant cultural role and reverence that the summit of Mauna Kea has always had within the indigenous Hawaiian community.  We are most fortunate and grateful to have the opportunity to conduct observations from this mountain.


\end{document}